\pdfoutput=1
\RequirePackage{ifpdf}
\ifpdf 
\documentclass[pdftex]{sigma}
\else
\documentclass{sigma}
\fi

\numberwithin{equation}{section}

\def\bla#1{$(${\it #1\/{}}$)$}

\def\DJ{\leavevmode\lower.6ex\hbox to 0pt{\hskip-.23ex
    \accent"16\hss}D}
\def\dj{\leavevmode\hbox to 0pt{\hskip.2ex
    \accent"16\hss}d}

\begin{document}

\allowdisplaybreaks

\renewcommand{\PaperNumber}{040}

\FirstPageHeading

\ShortArticleName{Dunkl Operators as Covariant Derivatives in a~Quantum Principal Bundle}

\ArticleName{Dunkl Operators as Covariant Derivatives
\\
in a~Quantum Principal Bundle}

\Author{Micho {\DJ}UR{\DJ}EVICH~$^\dag$ and Stephen Bruce SONTZ~$^\ddag$}

\AuthorNameForHeading{M.~{\DJ}ur{\dj}evich and S.B.~Sontz}

\Address{$^\dag$~Instituto de Matem\'aticas, Universidad Nacional Aut\'onoma de M\'exico,
\\
\hphantom{$^\dag$}~Circuito Exterior, Ciudad Universitaria, CP 04510, Mexico City, Mexico}
\EmailD{\href{mailto:micho@matem.unam.mx}{micho@matem.unam.mx}}

\Address{$^\ddag$~Centro de Investigaci\'on en Matem\'aticas, A.C.\ (CIMAT),
\\
\hphantom{$^\ddag$}~Jalisco s/n, Mineral de Valenciana, CP 36240, Guanajuato, Gto., Mexico}
\EmailD{\href{mailto:sontz@cimat.mx}{sontz@cimat.mx}}

\ArticleDates{Received November 01, 2012, in f\/inal form May 17, 2013; Published online May 30, 2013}

\Abstract{A quantum principal bundle is constructed for every Coxeter group acting on
a~f\/inite-dimensional Euclidean space~$E$, and then a~connection is also def\/ined on this bundle.
The covariant derivatives associated to this connection are the Dunkl operators, originally introduced as
part of a~program to generalize harmonic analysis in Euclidean spaces.
This gives us a~new, geometric way of viewing the Dunkl operators.
In particular, we present a~new proof of the commutativity of these operators among themselves as
a~consequence of a~geometric property, namely, that the connection has curvature zero.}


\Classification{20F55; 81R50; 81R60}

\section{Introduction}

A major development in modern harmonic analysis is the generalization of the partial derivative operators
acting on functions on Euclidean space to the larger class of Dunkl operators.
This theory was introduced in 1989 by Dunkl in his paper~\cite{DU}.
See~\cite{JEU,DX,MR, RO1,MV} as well as references therein for further mathematical
developments, including generalizations of the Laplacian operator (known as the Dunkl Laplacian) and the
Fourier transform (known as the Dunkl transform).
This theory has had applications as well in mathematical physics, pro\-ba\-bility and algebra; these include
studies of Calogero--Moser--Sutherland and other integrable systems (see~\cite{ETI,opdam, SU}), Segal--Bargmann spaces of Dunkl type and their associated integral kernel transforms
(see~\cite{SBS}), relations to Hecke algebras (see~\cite{CH}) and rational Cherednik algebras
(see~\cite{ETI}) as well as Markov processes generalizing Brownian motion (see~\cite{MV}).
Especially good introductions are~\cite{opdam} and~\cite{MR}.

Also in the 1980's there arose interest in a~new type of geometry, which has usually been called
non-commutative geometry (see~\cite{CO,W-q}) but also called quantum geometry (see~\cite{Prug} for
a~dif\/ferent formalization of similar geometrical ideas), and along similar lines a~new type of group
theory known as quantum group theory (see~\cite{DR,W-cmpg}).
Following the approach of Woronowicz (see~\cite{W-cmpg,W-diff}), the f\/irst author has
developed an extensive theory of quantum principal \mbox{bundles} (see~\cite{D-qclass, D1, D2,D3}). This
theory includes connections on such bundles as well as their asso\-cia\-ted covariant derivatives and curvature.
In this paper this latter theory is applied to construct a~quantum principal bundle and a~connection on it
with the property that the asso\-ciated \mbox{covariant} derivatives are exactly the Dunkl operators.
Using this identif\/ication we provide a~new, geometric proof that the Dunkl operators commute among
themselves.
This proof is based on the fact, which we prove, that the curvature of the connection is zero.

The covariant derivatives of a~connection in classical dif\/ferential geometry are local operators.
Therefore Dunkl operators, being non-local when the multiplicity function is non-zero, can not be covariant
derivatives in the classical context.
So our results show the power of using non-commutative structures.
A way of describing our results in physics terminology is that the theory of Dunkl operators is a~type of
non-commutative gauge theory.

The paper is organized as follows.
In the next section we review the def\/initions and basic facts about root systems, Coxeter groups and
Dunkl operators.
We then have a~section with the results of this paper and their proofs.
See equation~\eqref{dkbx} for a~mathematical formula, one case of which expresses the title of this paper.
Two appendices discuss quantum principal bundles and the dif\/ferential calculi for f\/inite classical
groups, but even so the paper is not self-contained.

\section{Dunkl operators}

We now review some basic properties of Dunkl operators, which were introduced in~\cite{DU}.
Other references for this material are~\cite{GB,HU} and~\cite{MR}.
These may be consulted for more details and proofs.

We let $\mathbb{R}^n$ for $n\ge1$ denote the f\/inite-dimensional Euclidean space equipped with the
standard inner product $\left\langle x,y\right\rangle=x_1y_1+\dots+x_n y_n$ for $x=(x_1,\dots,x_n)$ and
$y=(y_1,\dots,y_n)\in\mathbb{R}^n$.
For any $x\in\mathbb{R}^n$ we use its standard Euclidean norm $||x||=\sqrt{\langle x,x\rangle}$.

We would like to emphasize the intrinsic geometrical nature of our constructions.
\mbox{Accordingly} we shall be working within a~given f\/inite-dimensional Euclidean vector space $E$, providing
a~framework for our considerations which are independent of a~particular choice of coordinates.
Being a~f\/inite-dimensional Euclidean space, $E$ is always isomorphic to one of the standard spaces
$\mathbb{R}^n$, where $n=\dim (E)$.
These isomorphisms are in a~natural correspondence with the orthonormal frames in $E$.

For any $0\ne\alpha\in E$ we denote by $\sigma_\alpha$ the orthogonal ref\/lection in the hyperplane
$H_{\alpha}$ orthogonal to $\alpha$, namely $H_{\alpha}=\{x\in E~|~\left\langle\alpha,x\right\rangle=0\}$.
Then we have the explicit formula
\begin{gather*}
(x)\sigma_\alpha\equiv x\sigma_\alpha=x-\frac{2\left\langle\alpha,x\right\rangle}{||\alpha||^2}\alpha
\end{gather*}
for all $x\in E$.
A simple calculation shows that $\langle x\sigma_\alpha,y\sigma_\alpha\rangle=\langle x,y\rangle$ holds for
all $x,y\in E$, that is, $\sigma_\alpha\in\mathrm{O}(E)$, the orthogonal group of $E$.
One can easily prove many elementary properties such as $\det(\sigma_\alpha)=-1$ and $\sigma_\alpha^2=I$,
the identity element in $\mathrm{O}(E)$.

The reason for writing $\sigma_\alpha$ on the right is that in geometry the action of a~group on the total
space of a~principal bundle is by convention a~right action.
\begin{definition}
\label{def-21}
A {\em root system} is a~f\/inite set ${R}$ of non-zero vectors in $E$ which satisf\/ies the following
properties:
\begin{enumerate}\itemsep=0pt
\item[1)] $\alpha\in{R}\Rightarrow-\alpha\in{R}$,

\item[2)] $\alpha\in{R}$ and $r\alpha\in{R}$ for some $r\in\mathbb{R}\Rightarrow r=\pm1$,

\item[3)] $({R})\sigma_\alpha={R}$ for all $\alpha\in{R}$,

\item[4)] $||\alpha||^2=2$ for each root $\alpha\in{R}$.
\end{enumerate}

\begin{remark}
Even though Property~3 implies Property~1 since $\alpha\sigma_\alpha=-\alpha$, redundancy does no harm.
Property~4 is a~standard normalization convention.
Since the roots $\alpha$ are only used to def\/ine the ref\/lections $\sigma_{\alpha}$ and
$\sigma_{r\alpha}=\sigma_{\alpha}$ for any $0\ne r\in\mathbb{R}$, this normalization does not inf\/luence
our results while it allows for compatibility with some other authors.
\end{remark}

Given such a~root system ${R}$, we denote the subgroup of $\mathrm{O}(E)$ generated by the elements
$\sigma_\alpha$ for all $\alpha\in{R}$ as $G\equiv G({R})$.
(It turns out that this group is f\/inite, since it is isomorphic to a~subgroup of the group of
permutations of ${R}$, which itself is a~f\/inite group.
See~\cite{HU} for a~proof.) We say that $G$ is the {\em $($finite$)$ Coxeter group} associated with the root
system ${R}$.

A $G$-invariant function $\varkappa:{R}\rightarrow\mathbb{C}$ is called a~{\em multiplicity function}.
(Note that we are using Property~3 above of a~root system here, since we are requiring that
$\varkappa_{\alpha g}=\varkappa_\alpha$ for $g\in G$ and $\alpha\in{R}$ and so we need to know that $\alpha
g\in{R}$, the domain of $\varkappa$.)
\end{definition}

For a~f\/ixed root system ${R}$ the set of all multiplicity functions def\/ined on ${R}$ forms a~f\/inite-dimensional vector space over $\mathbb{C}$ of dimension equal to the number of $G$-orbits in ${R}$.

We note that according to our def\/inition ${R}=\varnothing$, the empty set, is a~root system whose
associated group consists of exactly one element (the identity), and therefore the trivial subgroup of
$\mathrm{O}(E)$ is a~Coxeter group.
Non-trivial examples of root systems and their associated Coxeter groups are given in~\cite{MR}.
See the text~\cite{HU} for much more information on Coxeter groups.

We let ${R}^+$ denote the subset of positive elements in ${R}$ with respect to a~given total order on~$E$.
(An \textit{order} on~$E$ is a~partial order $<$ such that $u<v\Rightarrow u+w<v+w$ and $ru<rv$ for all
$u,v,w\in E$ and $r>0$.
Such an order is said to be \textit{total} if for all $u,v\in E$ either $u<v$ or $v<u$ or $u=v$. 
We say $u\in E$ is \textit{positive} if $0<u$.) We def\/ine the Dunkl operators below in terms of the
subset ${R}^+$ of positive elements with respect to a~given total order (which {\em do} exist), since this
is how it is usually done in the literature.
However, this does not depend on the particular choice of total order.
This is due to various basic facts the reader can verify such as ${R}={R}^+\cup(-{R}^+)$ (a disjoint
union), $\mathbb{R}\alpha\cap{R}=\{\alpha,-\alpha\}$ for all $\alpha\in{R}$,
$\sigma_{-\alpha}=\sigma_{\alpha}$, $\alpha\sigma_\alpha=-\alpha$ and
$\varkappa_{-\alpha}=\varkappa_\alpha$.
\begin{definition}
\label{define-dunkl-operator}
For any $\xi\in E$ and multiplicity function $\varkappa:{R}\to\mathbb{C}$ we def\/ine the {\em Dunkl
\mbox{operator}}~$T_{\xi,\varkappa}$ by
\begin{gather*}
T_{\xi,\varkappa}f(x):=\partial_\xi f(x)+\sum_{\alpha\in{R}^+}
\varkappa_\alpha\frac{\langle\alpha,\xi\rangle}{\langle\alpha,x\rangle}\big(f(x)-f(x\sigma_\alpha)\big),
\end{gather*}
where $\partial_\xi=\langle\xi,{\rm grad}\rangle$ is the directional derivative associated to $\xi$ (with
${\rm grad}=(\partial/\partial x_1,\dots,\partial/\partial x_n)$ being the usual gradient operator in
orthogonal coordinates), $x\in E$ and $f\in C^1(E)$ is a~complex valued function.
This def\/inition can equivalently be written as
\begin{gather*}
T_{\xi,\varkappa}f(x)=\partial_\xi f(x)+\dfrac{1}{2}\sum_{\alpha\in{R}}
\varkappa_\alpha\frac{\langle\alpha,\xi\rangle}{\langle\alpha,x\rangle}\big(f(x)-f(x\sigma_\alpha)\big),
\end{gather*}
which shows that this operator does not depend on the choice of the total order.
\end{definition}

Note that the linear operator $T_{\xi,\varkappa}$ depends linearly on $\xi$.
For the constant multiplicity function $\varkappa\equiv0$ or for ${R}=\varnothing$ (either case being called
the \textit{trivial Dunkl structure}) the opera\-tor~$T_{\xi,\varkappa}$ reduces to the directional
derivative $\partial_\xi$ associated to the vector $\xi\in E$.

Of course this def\/inition of $T_{\xi,\varkappa}f(x)$ only makes sense for $x\notin\cup_\alpha H_\alpha$,
where $H_{\alpha}$ is the hyperplane def\/ined above.
However, we have the identity
\begin{gather*}
\dfrac{f(x)-f(x\sigma_\alpha)}{\left\langle\alpha,x\right\rangle}=\int_0^1\mathrm{d}t\,\,\partial_{\alpha}
f\big(t x+(1-t)x\sigma_\alpha\big)
\end{gather*}
for all $x\notin H_\alpha$.
This allows one to use the expression on the right side to def\/ine the Dunkl operator at every $x\in E$,
provided that $f$ is a~$C^1$ function so that the directional derivative~$\partial_{\alpha}f$ is def\/ined
and continuous.
Specif\/ically for $x\in H_\alpha$ (that is, $x\sigma_{\alpha}=x$) the right side of the previous equation
becomes $\partial_{\alpha}f(x)$.
The Dunkl operators also have realizations in other function spaces, most notably as anti-Hermitian
operators acting in $L^2(E,w_\varkappa \mathrm{d}^n x)$, where $\mathrm{d}^n x$ denotes Lebesgue measure
on $E$ and $w_\varkappa$ is a~non-negative function def\/ined on $E$.

For us one important result is the generalization of the Leibniz rule to this context.
Since this result is not emphasized in the literature (though it appears in Proposition~4.4.12 in~\cite{DX} for polynomials), we now present the statement and proof.
\begin{theorem}
\label{Leibniz-F}
For any $f,g:E\to\mathbb{C}$ which are $C^1$ and for any $x\in E$ we have the following generalization of
the Leibniz formula:
\begin{gather*}
(T_{\xi,\varkappa}(f g))(x)=(T_{\xi,\varkappa}f)(x)g(x)+f(x)(T_{\xi,\varkappa}g)(x)
\\
\phantom{(T_{\xi,\varkappa}(f g))(x)=}
{} -\sum_{\alpha\in{R}^+}\varkappa_\alpha\dfrac{\left\langle\alpha,\xi\right\rangle}
{\left\langle\alpha,x\right\rangle}(f(x)-f(x\sigma_\alpha))(g(x)-g(x\sigma_\alpha)).
\end{gather*}
\end{theorem}

\begin{proof}
We prove the result for all $x\notin\cup_\alpha H_\alpha$ and use the previous comments to extend it to all
$x\in E$.
We use the fact that directional derivatives satisfy the usual Leibniz rule and proceed as follows:
\begin{gather*}
(T_{\xi,\varkappa}(f g))(x)-(T_{\xi,\varkappa}f)(x)g(x)-f(x)(T_{\xi,\varkappa}g)(x)
\\
=\sum_{\alpha\in{R}^+}\varkappa_\alpha\dfrac{\left\langle\alpha,\xi\right\rangle}
{\left\langle\alpha,x\right\rangle}\Big(f(x)g(x)\!-\!f(x\sigma_{\alpha})g(x\sigma_{\alpha}
)\!-\!(\,f(x)\!-\!f(x\sigma_{\alpha})\,)g(x)\!-\!f(x)(\,g(x)\!-\!g(x\sigma_{\alpha}) )\Big)        
\\
=\sum_{\alpha\in{R}^+}\varkappa_\alpha\dfrac{\left\langle\alpha,\xi\right\rangle}
{\left\langle\alpha,x\right\rangle}\Big({-}f(x)g(x)+f(x\sigma_{\alpha})g(x)+f(x)g(x\sigma_{\alpha}
)-f(x\sigma_{\alpha})g(x\sigma_{\alpha})\Big)
\\
=-\sum_{\alpha\in{R}^+}\varkappa_\alpha\dfrac{\left\langle\alpha,\xi\right\rangle}
{\left\langle\alpha,x\right\rangle}(f(x)-f(x\sigma_\alpha))(g(x)-g(x\sigma_\alpha)).
\end{gather*}
And so the theorem is proved.
\end{proof}

We have an immediate consequence of this theorem.
\begin{corollary}
Suppose that $f$ and $g$ are as in the previous theorem.
Suppose that at least one of these functions is $G$-invariant.
Then we have the usual Leibniz formula
\begin{gather*}
(T_{\xi,\varkappa}(f g))(x)=(T_{\xi,\varkappa}f)(x)g(x)+f(x)(T_{\xi,\varkappa}g)(x)
\end{gather*}
for all $x\in E$.
\end{corollary}
A non-trivial result of this theory is that the operators $T_{\xi,\kappa}$ and $T_{\eta,\kappa}$ commute
for all $\xi,\eta\in E$.
This fundamental result was f\/irst proved in Dunkl's seminal article~\cite{DU}.
Another proof based on the Koszul complex is given in~\cite{DJO}.
We of\/fer a~new non-commutative geometric proof later on.

\section{Results}
\label{results}

We shall now apply the general formalism of connections and quantum principal bundles to the special case
when the structure group $G$ is a~f\/inite Coxeter group associated to a~root system ${R}$ in
a~f\/inite-dimensional Euclidean vector space $E$.
We shall derive some of the basic expressions and properties for Dunkl operators as consequences of
geometrical conditions involving connections and their covariant derivatives and curvature.
The reader should consult the appendices for all new notation.
We also use, often without comment, Sweedler's elegant notation for coproducts and co-actions to avoid an
excessive amount of summations and indices.

Being a~Coxeter group, $G$ possesses a~distinguished dif\/ferential calculus $\Gamma$.
It is based on the set $S$ of all ref\/lections of $G$.
A property of Coxeter groups is that every ref\/lection $s\in G$ has the form $s=\sigma_\alpha$ where
$\alpha\in{R}$ is a~root vector (see~\cite{HU}). Since the conjugation by any element in $G$ of a~ref\/lection is again a~ref\/lection and
since the ref\/lections are involutive, the associated calculus $\Gamma$ turns out to be bicovariant and
$*$-covariant.
The elements $\pi(s)=[s]$ where $s\in S$ form a~canonical basis in the complex vector space of
left-invariant elements $\Gamma_{{\rm inv}}$ (cf.\ Appendix~\ref{appendixB}).

We def\/ine the total space of a~principal bundle by
\begin{gather*}
P:=E\setminus(\cup_{\alpha\in{R}}H_\alpha),
\end{gather*}
which is an open, dense subset of $E$.
The natural right action of $G$ on $E$ (given in the previous section) leaves $P$ invariant.
Moreover, the restricted action of $G$ on $P$ is free, since $P$ consists exactly of all the vectors in $E$
having trivial stabilizers in $G$.
In this way we have def\/ined a~classical principal bundle $P\to P/G$.
The space $P/G$ is dif\/feomorphic to any of the connected components of $P$.
It is called an open Weyl chamber.
So $P$ is equipped with the classical dif\/ferential calculus, based on smooth dif\/ferential forms.
However, these forms will be interpreted as {\it hori\-zon\-tal forms} on $P$.
The full dif\/ferential calculus $\Omega(P)$ will be quantum and constructed as explained in Appendix~\ref{appendixA}.
We wish to emphasize that even though the spaces~($G$,~$P$ and $P/G$) of this principal bundle are all
classical, the dif\/ferential calculus which we will use is quantum and not classical.

By def\/inition of the Coxeter group associated to a~root system, we have $g^{-1}\sigma_\alpha
g=\sigma_{\alpha g}$, where we always use the convention that $G$ acts on the right in $E$.

Let $\varpi$ be the canonical f\/lat connection, as def\/ined in Appendix~\ref{appendixA}.
Arbitrary connections $\omega$
are then given by connection displacement maps $\lambda\colon\Gamma_{{\rm
inv}}\rightarrow\mathfrak{hor}^1(P)$ so that $\omega=\varpi+\lambda$.
For more details and def\/initions, see Appendix~\ref{appendixA}.
Such maps $\lambda$ constitute the vector space
associated to the af\/f\/ine space of all connections on~$P$, and satisfy two characterizing conditions:
hermicity and covariance.
We shall now analyze possible forms for the connection displacement maps~$\lambda$.

If we def\/ine $\lambda$ by
\begin{gather}
\label{generalized-dunkl}
\lambda[\sigma_\alpha](x)=ih_\alpha(x)\alpha
\end{gather}
for all root vectors $\alpha\in R$, where $h_\alpha$ are smooth real functions on $P$ satisfying
\begin{gather*}
h_{\alpha g}(x)=h_\alpha\big(x g^{-1}\big),
\qquad
h_{-\alpha}(x)=h_\alpha(x),
\end{gather*}
then such a~map $\lambda$ will be a~connection displacement.
Here the root vectors $\alpha$ are interpreted in a~natural way as one-forms on~$P$.

\begin{remark}
The presence of the imaginary unit $i=\sqrt{-1}$ in the above formula~\eqref{generalized-dunkl} is due to
the fact that we are considering real connections which intertwine the corresponding $*$-structures.
It is worth mentioning that $[\sigma_\alpha]^{*}=-[\sigma_\alpha]$, that is, the generators of our canonical
dif\/ferential calculus are imaginary.
\end{remark}

The f\/irst condition ensures the covariance of $\lambda$ while the second condition is a~necessary
consistency requirement, because the value of $\lambda$ on $[\sigma_\alpha]$ obviously should not change if
we replace $\alpha$ by $-\alpha$.
(Recall that $\sigma_{-\alpha}=\sigma_{\alpha}$.) Let us explicitly verify the covariance property.
It reads
\begin{gather*}
ih_\alpha(x g)\alpha g^{-1}=\bigl(\lambda[\sigma_\alpha]\bigr)_{g}(x)=\lambda\big[g\sigma_{\alpha}g^{-1}
\big](x)=\lambda\big[\sigma_{\alpha g^{-1}}\big](x)=ih_{\alpha g^{-1}}(x)\alpha g^{-1},
\end{gather*}
and this is equivalent to the covariance condition for the functions $h_\alpha$.
We require these functions to be real valued, so that the resulting connection $\omega$ is real.
\begin{definition}
Connections $\omega=\varpi+\lambda$, where $\lambda$ is given by~\eqref{generalized-dunkl}, will be called
{\it Dunkl connections}.
\end{definition}

Let us also observe that using the same type of functions $h_\alpha$ we can generate another class of
`spherically symmetrical' displacements, where
\begin{gather*}
\lambda[\sigma_\alpha](x)=ih_\alpha(x)\zeta(x),
\end{gather*}
where $\zeta$ is the canonical radial one-form given by
\begin{gather*}
\zeta(x)=\frac{1}{2}D\langle x,x\rangle
\end{gather*}
and $D\colon\mathfrak{hor}(P)\rightarrow\mathfrak{hor}(P)$ is the standard de Rham derivative of classical
dif\/ferential forms on~$P$.

\begin{proposition}
With $D$ as above, the covariant derivative associated to the Dunkl connection $\omega=\varpi+\lambda$ is
given by
\begin{gather*}
D_\omega(\varphi)(x)=D(\varphi)(x)+i\sum_{\alpha\in{R}^+}
h_\alpha(x)\bigl(\varphi(x)-\varphi(x\sigma_\alpha)\bigr)\alpha.
\end{gather*}
\end{proposition}
\begin{proof}
This follows directly from the general expression for covariant derivatives of connections as presented in
the appendices.
\end{proof}

We can now use the following ansatz to generate an interesting class of Dunkl connections:
\begin{gather}
\label{generic-lambda}
h_\alpha(x)=\psi_{\alpha}[\langle\alpha,x\rangle],
\end{gather}
where $\psi_\alpha\colon\mathbb{R}\setminus\{0\}\to\mathbb{R}$ are smooth, {\it odd} functions indexed by
the elements of $\alpha\in{R}$ in a~$G$-invariant way (so ef\/fectively they are indexed by orbits of the
action of $G$ on ${R}$).

Note that for all $\alpha\in{R}$ the scalar products $\langle x,\alpha\rangle$ are never zero, because
$x\in P$.
As a~very particular case of the above expression, we can consider for $r\in\mathbb{R}\setminus\{0\}$
\begin{gather}
\label{standard_dunkl}
\psi_{\alpha}(r)=\varkappa_\alpha\frac{1}{r},
\end{gather}
where $\varkappa\colon{R}\rightarrow\mathbb{C}$ is a~(necessarily $G$-invariant and, for the purpose of
having a~real connection, real-valued) multiplicity function def\/ined on ${R}$.
With such a~choice we will be able to reproduce the standard expression for the Dunkl operators as given in
the previous section.
\begin{definition}
\label{definition_standard_dunkl}
We call the connection associated to the choice~\eqref{standard_dunkl} for $\psi_\alpha$ the
\textit{standard Dunkl connection}.
\end{definition}
Indeed, the formula for the covariant derivative gives in this case
\begin{gather*}
D_\omega(\varphi)(x)=D(\varphi)(x)+i\sum_{\alpha\in{R}^+}
\varkappa_\alpha\frac{\varphi(x)-\varphi(x\sigma_\alpha)}{\langle x,\alpha\rangle}\alpha,
\end{gather*}
which for functions corresponds to the classical def\/inition up to a~factor of $i$.
\begin{remark}
This factor of $i$ corresponds to our geometric condition of reality of the connection.
However, from the point of view of the algebraic considerations we develop here this is inessential.
See the concluding section for more details.
\end{remark}

It is worth mentioning that we can also use the more general ansatz
\begin{gather}
\label{generic-lambda-2}
h_\alpha(x)=\psi_{\alpha}[\langle x,\alpha\rangle,\langle x,x\rangle],
\end{gather}
where now $\psi_\alpha$ are smooth functions of two variables and the indexing by the $\alpha$'s is done as
before in a~$G$-invariant way.

We can consider situations where there is no indexing at all in the sense that
\begin{gather*}
\psi_\alpha(u,v)=\psi(u,v)
\end{gather*}
for a~single smooth function $\psi$.
It turns out that in this case the ansatz~\eqref{generic-lambda-2} captures all `generic' displacements.
By generic displacements, we understand those that are naturally associated, by using~\eqref{gen-disp-def}
below, to $\mathrm{O}(E)$-covariant maps $\mu\colon\mathrm{RP}(E)\rightarrow\Lambda^1(E)$ def\/ined on the
real projective space $\mathrm{RP}(E)$ associated to $E$.
Also $\Lambda(E)$ is the algebra of classical complex dif\/ferential forms on $E$.
In other words, we assume that
\begin{gather}
\label{gen-disp-cov}
\mu\big(g p g^{-1}\big)=\mu(p)_g
\end{gather}
for every $g\in\mathrm{O}(E)$ and $p\in\mathrm{RP}(E)$.
The elements of $\mathrm{RP}(E)$ are interpreted as orthogonal projections onto the one-dimensional
subspaces of $E$.
Generic displacements $\lambda$ are characterized by
\begin{gather}
\label{gen-disp-def}
\lambda[\sigma_\alpha]=\mu(p_\alpha),
\end{gather}
where $p_\alpha=(1-\sigma_\alpha)/2$ is the projection onto the space $\mathbb{R}\alpha$.
Without loss of generality, we can assume that all vectors $\alpha$ are normalized as in
Def\/inition~\ref{def-21}, part~4.
\begin{proposition}\quad

\begin{enumerate}\itemsep=0pt
\item[$(i)$] If $\lambda$ is a~generic displacement $($in particular, real in the sense of intertwining the
corresponding $*$-structures$)$, then for any given $\alpha\in{R}$ we have that $\xi=-i\lambda[\sigma_\alpha]$
is a~real one-form, invariant under $\sigma_\alpha$ and also invariant under the group of all orthogonal
transformations that have $\alpha$ as a~f\/ixed point.
$($Let us call this group $\mathrm{O}(\alpha,E).)$

\item[$(ii)$] Conversely, if $\xi=\xi^{*}$ is a~one-form on $E$ invariant under $\sigma_\alpha$ and
under all transformations from $\mathrm{O}(\alpha,E)$, then there exists a~unique displacement $\lambda$
such that $\lambda[\sigma_\alpha]=i\xi$.

\item[$(iii)$] Every such one-form can be naturally decomposed $($in the region outside of the line
$\mathbb{R}\alpha)$ as
\begin{gather}
\label{O-alpha-E-inv}
\xi(x)=\psi\bigl\{\langle x,\alpha\rangle,\langle x,x\rangle\bigr\}\alpha+\tilde{\psi}
\bigl\{\langle x,\alpha\rangle,\langle x,x\rangle\bigr\}\zeta,
\end{gather}
where $\psi,\tilde{\psi}\colon\mathbb{R}\times\mathbb{R^+}\rightarrow\mathbb{R}$ are smooth functions
satisfying
\begin{gather}
\label{sigma-alpha-invariance}
\psi(u,v)=-\psi(-u,v)
\qquad
\text{and}
\qquad
\tilde{\psi}(u,v)=\tilde{\psi}(-u,v)
\end{gather}
and $\zeta$ is the canonical radial one-form on $E$.
Moreover, the displacement $\lambda$ will be closed in the standard sense of $D\lambda=0$ if and only if
\begin{gather}
\label{D-xi}
\frac{\partial\psi}{\partial v}=-\frac{\partial\tilde{\psi}}{\partial u}.
\end{gather}
\end{enumerate}
\end{proposition}

\begin{proof}
If $\lambda$ is a~generic displacement, then the covariance formula~\eqref{gen-disp-cov} implies that the
ref\/lection $\sigma_\alpha$ as well as the group $\mathrm{O}(\alpha,E)$ act trivially on
$\xi=-i\lambda[\sigma_\alpha]$.
The elements $\xi$ are real one-forms because $\lambda$, being a~connection displacement, commutes with the
corresponding $*$-structures (and the $[\sigma_\alpha]$ are {\it imaginary} in $\Gamma_{{\rm inv}}$).
And conversely, if we f\/ix $\alpha\in R$ and a~real $\sigma_\alpha$-invariant and
$\mathrm{O}(\alpha,E)$-invariant one-form $\xi$ on $E$, then every generic displacement $\lambda$
satisfying
\begin{gather*}
\lambda[\sigma_\alpha]=\mu(p_\alpha)=-i\xi
\end{gather*}
should also satisfy
\begin{gather*}
\lambda[\sigma_{\alpha g^{-1}}]=\mu(p_{\alpha g^{-1}})=\mu\big(g p_\alpha g^{-1}\big)=\mu(p_\alpha)_g=-i\xi_g,
\end{gather*}
where $g\in\mathrm{O}(E)$ is such that $\alpha g^{-1}\in R$.
This shows that such a~displacement is unique, if it exists.
(The orthogonal group acts transitively on $\mathrm{RP}(E)$ and in particular we can obtain all root
vectors by acting on $\alpha$ via appropriate orthogonal transformations).
What remains to prove is that the above formula consistently def\/ines the displacement $\lambda$.
If $h\in\mathrm{O}(E)$ is another element such that $\alpha h^{-1}=\pm\alpha g^{-1}$ then either
$g^{-1}h\in\mathrm{O}(\alpha,E)$ or $g^{-1}h\in\sigma_\alpha\mathrm{O}(\alpha,E)$.
Therefore
\begin{gather*}
\xi_h=\xi_{g(g^{-1}h)}=(\xi_{g^{-1}h})_g=\xi_g,
\end{gather*}
which shows the consistency of the def\/inition formula for $\lambda$.
The formula~\eqref{O-alpha-E-inv} is a~generic expression for $\mathrm{O}(\alpha,E)$-invariant one-forms.
Indeed, decomposing $E$ as
\begin{gather*}
E=\mathbb{R}\alpha\oplus(\mathbb{R}\alpha)^\bot
\end{gather*}
we see that $\xi$ must be an appropriate combination of $\zeta-(x,\alpha)\alpha/2$ which is the canonical
radial form in $\mathbb{R}\alpha^\bot$, and $\alpha$ which corresponds to the f\/irst summand
$\mathbb{R}\alpha$ in the above decomposition.
In other words, we have
\begin{gather*}
\xi(x)=q(x)\alpha+\tilde{q}(x)\zeta,
\end{gather*}
where $q$ and $\tilde{q}$ are smooth functions on $E\setminus\mathbb{R}\alpha$.
(We excluded the line $\mathbb{R}\alpha$ where the forms~$\alpha$ and~$x$ are proportional in order to have
a~unique factorization.)
In order to preserve the $\mathrm{O}(\alpha,E)$-symmetry the functions $q$ and $\tilde{q}$ can only depend
on the f\/irst coordinate as well as the radial part of the second coordinate in the above decomposition.
To put it in equivalent terms, we obtain
\begin{gather*}
q(x)=\psi\bigl(\langle x,\alpha\rangle,\langle x,x\rangle\bigr),
\qquad
\tilde{q}(x)=\tilde{\psi}\bigl(\langle x,\alpha\rangle,\langle x,x\rangle\bigr).
\end{gather*}
Equations~\eqref{sigma-alpha-invariance} are equivalent to the $\sigma_\alpha$-invariance of $\xi$.
Indeed, the radial one-form $\zeta$ is $\sigma_\alpha$-invariant, while $\alpha$ changes sign under the
ref\/lection by $\sigma_\alpha$.
Finally~\eqref{D-xi} is a~matter of direct calculation of the dif\/ferential, taking into account $D\langle
x,\alpha\rangle=\alpha$ and $D\langle x,x\rangle=2\zeta$.
\end{proof}
\begin{remark}
Due to covariance we see that all generic displacements are of the form
\begin{gather*}
\lambda[\sigma_\alpha](x)=i\psi\bigl(\bigl\langle\alpha,x\rangle,\langle x,x\rangle\bigr)\alpha+i\tilde{\psi}\bigl(\langle\alpha,x\rangle,\langle x,x\rangle\bigr)x,
\end{gather*}
where $\alpha\in R$ is now an arbitrary root vector.
\end{remark}

So we can get~\eqref{generic-lambda} as an important special case (with a~constant weight function, since
all root vectors are equivalent in this picture) when $\psi$ does not depend on the radial variable $v$ and
$\tilde{\psi}=0$.
Another special case is when $\psi=0$ and $\tilde{\psi}$ does not depend on $u$.
In the latter case we get a~spherically symmetrical displacement $\lambda$ taking the same value on all of
the genera\-tors~$[\sigma_\alpha]$.

Of course, an arbitrary displacement need not be generic.
With the help of local trivializations of the bundle $P$, as indicated in Appendix~\ref{appendixB},
we can easily classify all possible displacements, in terms of their local representations.

Here is a~version of the above proposition for arbitrary displacements.
The only symmetries we have are those of the Coxeter group~$G$, and in general this group does not act
transitively on~$R$.
For each~$\alpha\in{R}$ we let~$G_\alpha$ be the stabilizer of $\alpha$~in~$G$, and we let~$O_\alpha$ be
the orbit of~$\alpha$ under the action of~$G$.
Clearly $O_{-\alpha}=O_\alpha$ and we have a~natural identif\/ication $G/G_\alpha\leftrightarrow O_\alpha$
between~$O_\alpha$ and the right cosets of~$G_\alpha$ in $G$.
\begin{proposition}\quad
\begin{enumerate}\itemsep=0pt
\item[$(i)$] If $\lambda$ is an arbitrary displacement, then $\xi_\alpha=-i\lambda[\sigma_\alpha]$ is real and
invariant under~$\sigma_\alpha$ as well as under all of the elements in $G_\alpha$.

\item[$(ii)$] Conversely, let us choose for every orbit $O$ of the action of $G$ in $R$ one element
$\alpha\in O$.
$($So that we have $O=O_\alpha.)$ Let us assume that for every such representative $\alpha$ a~real one-form~$\xi_\alpha$ on $P$ is given which is invariant under~$\sigma_\alpha$ and under all of the transformations
in~$G_\alpha$.
Then there exists a~unique displacement $\lambda$ such that $\lambda[\sigma_\alpha]=i\xi_\alpha$ for every
representing root vector $\alpha$.

\item[$(iii)$]  Moreover, the displacement $\lambda$ will be closed if and only if
\begin{gather*}
\xi_\alpha=D\psi_\alpha
\end{gather*}
for each $\alpha\in R$, where $\psi_\alpha$ are smooth real functions on $P$ invariant under $G_\alpha$ and
under~$\sigma_\alpha$.
We can naturally make the system of functions $ \{\psi_\alpha \}_{\alpha\in{R}}$ covariant in the
sense that
\begin{gather*}
\psi_{\alpha g}(x)=\psi_\alpha\big(xg^{-1}\big)
\end{gather*}
for every $x\in P$ and $g\in G$.
\end{enumerate}
\end{proposition}

\begin{proof}
To prove $(i)$ and $(ii)$ we can follow conceptually the same line of reasoning as for the
corresponding properties of the previous proposition.
If the forms $\lambda[\sigma_\alpha]$ are closed, then due to the contractibility of the space~$P$, they
are globally exact.
It is worth mentioning that due to the covariance of~$\lambda$ it is suf\/f\/icient to check closedness for
one representative of each orbit~$O$.
Finally, the possibility of choosing a~$G$-covariant system $\{\psi_\alpha\}_{\alpha\in R}$ is a~direct
consequence of the $G$-covariance of the dif\/ferential~$D$ and the fact that a~short exact sequence of
$G$-modules always splits.
(This fact holds for arbitrary compact groups, whether classical or quantum.)
\end{proof}

\begin{remark}
So we have a~kind of {\it elementary} displacement where we pick out a~single orbit $O$ and one element
$\alpha\in O=O_\alpha$ and then proceed by constructing a~one-form $\xi_\alpha$.
For the representatives of the other orbits we can put $\xi_\alpha=0$.
In other words the associated displacements will vanish on all elements $[\sigma_\beta]$ if $\beta\not\in
O$.
\end{remark}

Let us now calculate the curvature of the displaced connection $\omega=\varpi+\lambda$.
Let us recall that, according to the general formalism and as explained in Appendix~\ref{appendixA},
this curvature is related to the original curvature via
\begin{gather}
r_{\omega}(a)=r_\varpi(a)+D_\varpi\lambda\pi(a)+\lambda\pi\big(a^{(1)}\big)\lambda\pi\big(a^{(2)}\big)
+\ell_\varpi\bigl(\pi\big(a^{(1)}\big),\lambda\pi\big(a^{(2)}\big)\bigr)
\nonumber
\\
\phantom{r_{\omega}(a)}{}
=D\lambda\pi(a)+\lambda\pi\big(a^{(1)}\big)\lambda\pi\big(a^{(2)}\big).
\label{curvature-displacement}
\end{gather}
Here we are using Sweedler's notation for the coproduct $\phi$ in the Hopf algebra $\mathcal{A}$, namely we
write $\phi(a)=a^{(1)}\otimes a^{(2)}$ instead of the more cumbersome
$\phi(a)=\sum\limits_{j=1}^{n_a}a^\prime_j\otimes a^{\prime\prime}_j$.

In our case, the initial connection $\varpi$ is the canonical f\/lat connection, so we have $D_\varpi=D$
and $r_\varpi=0$.
By construction, the canonical f\/lat connection is {\it regular} so we also have $\ell_\varpi=0$.

Our main goal now is to prove that under certain general conditions, the curvature of the displaced
connection vanishes too.
These general conditions will include as a~special case the classical Dunkl operators.
This means that it suf\/f\/ices to prove that the remaining terms in the above
expression~\eqref{curvature-displacement} vanish identically for appropriate displacements $\lambda$.
We shall also derive an explicit formula for calculating the curvature of a~general Dunkl connection (where
the third term is the only one that possibly does not vanish).
\begin{lemma}
For a~Dunkl connection given by
\begin{gather*}
\lambda[\sigma_\alpha](x)=i\psi_\alpha[\langle x,\alpha\rangle]\alpha
\end{gather*}
we have $D\lambda=0$.
In other words, the displacement $\lambda$ is closed as a~standard differential form on~$P$.
\end{lemma}

This is just a~convenient rephrasing of the previously established result.
It is worth remembering that in our quantum dif\/ferential calculus, classical dif\/ferential forms on $P$
play the role of horizontal dif\/ferential forms for the complete calculus on the bundle $P$.

So the curvature of such a~Dunkl connection is given by the third term in the expression for the displaced
curvature.
\begin{definition}
An element $\rho$ of the Coxeter group $G$ is called a~{\it $2$-rotation} if $\rho=\sigma_\alpha\sigma_\beta$
for some $\alpha,\beta\in{R}$ and $\alpha\neq\pm\beta$.
Such a~rotation is called {\it proper} if its period is greater than two (in other words $\alpha$ and
$\beta$ are not orthogonal).
\end{definition}

For a~given proper 2-rotation $\rho$ we can introduce in a~canonical way a~horizontal 2-form $w_\rho$ which
we can interpret as the volume element of the plane of the rotation of $\rho$.
Indeed if we def\/ine
\begin{gather*}
w_\rho=\alpha\wedge\beta,
\qquad
\mathrm{where}
\qquad
\rho=\sigma_\alpha\sigma_\beta,
\end{gather*}
then this formula determines $w_\rho$ up to a~sign.
We can f\/ix the ambiguity of the sign by requiring that $\alpha$ and $\beta$ follow the orientation of the
plane induced by $\rho$.
It is worth observing that $w_{\rho^{-1}}=-w_{\rho}.$
\begin{proposition}\quad
\begin{enumerate}\itemsep=0pt
\item[$(i)$] Let $G$ be a~Coxeter group.
Then the curvature $r_\omega$ of any connection $\omega=\varpi+\lambda$ vanishes on all elements of $G$
except possibly some proper two-dimensional rotations $\rho$ in which case we have
\begin{gather*}
r_\omega(\rho)={\sum}^{*}\lambda[\sigma_\alpha]\lambda[\sigma_\beta].
\end{gather*}
Here the sum ${\sum}^{*}$ is taken over all possible decompositions
$\rho=\sigma_\alpha\sigma_\beta$ into the product of two reflections in $G$.

\item[$(ii)$] Moreover, if $\omega$ is a~Dunkl connection, then we have
\begin{gather}
\label{dunkl-curvature}
r_\omega(\rho)=-w_\rho {\sum}^{*}h_\alpha h_\beta,
\end{gather}
where the root vectors participating in the above sum are such that $\alpha\wedge\beta=w_\rho$.
\end{enumerate}
\end{proposition}

\begin{proof}
Let us begin our proof by recalling from Appendix~\ref{appendixB}
that for each $g\in G$
\begin{gather*}
\phi(g)=g^{(1)}\otimes g^{(2)}=\sum_{h\in G}h\otimes\big(h^{-1}g\big),
\end{gather*}
where we are using Sweedler's notation in the second expression.
On the other hand, by the construction of the dif\/ferential calculus $\Gamma$ over $G$, the map $\pi$
vanishes on all elements except the ref\/lections $s=\sigma_\alpha$ and the neutral element $\epsilon$,
where we have $[\epsilon]=- \sum\limits_{s\in S}[s]$.

So the expression
\begin{gather*}
\delta(g)=(\pi\otimes\pi)\phi(g)=\pi\big(g^{(1)}\big)\otimes\pi\big(g^{(2)}\big)
\end{gather*}
vanishes on all $g\in G$ except possibly for the following three types of elements: The neutral element
$\epsilon\in G$, in which case
\begin{gather*}
\delta(\epsilon)=[\epsilon]\otimes[\epsilon]+\sum_{s\in S}[s]\otimes[s],
\end{gather*}
the ref\/lections $s\in S$, where we have
\begin{gather*}
\delta(s)=[\epsilon]\otimes[s]+[s]\otimes[\epsilon],
\end{gather*}
and f\/inally
\begin{gather*}
\delta(\rho)= {\sum}^{*}[\sigma_\alpha]\otimes[\sigma_\beta],
\end{gather*}
where $\rho$ is 2-rotation and the sum is taken over all decompositions $\sigma_\alpha\sigma_\beta=\rho$.

The f\/irst two types of expressions are symmetric elements of $\Gamma_{{\rm inv}}\otimes\Gamma_{{\rm
inv}}$ and hence, after being coupled with the classical one-form $\lambda$ and multiplication in the
exterior algebra of $P$, they vanish identically.
Regarding the third and the most interesting type of potentially non-zero expressions, namely those
associated with the 2-rotations, let us observe that if $\rho$ is involutive (in other words given by the
composition of two ref\/lections associated to mutually orthogonal root vectors $\alpha$ and $\beta$), then
$\delta(\rho)$ will also be symmetric because the term $[\sigma_\alpha]\otimes[\sigma_\beta]$ will appear
in the sum if and only if its pair $[\sigma_\beta]\otimes[\sigma_\alpha]$ also appears, since the
ref\/lections $\sigma_\alpha$ and $\sigma_\beta$ commute in this case.

So now the possibly non-zero terms in the formula for the curvature can only come from {\it proper}
2-rotations $\rho$.
The curvature is therefore explicitly given by the third term of the
decomposition~\eqref{curvature-displacement}, namely
\begin{gather*}
r_\omega[\rho]= {\sum}^{*}\lambda[\sigma_\alpha]\lambda[\sigma_\beta],
\qquad
\mathrm{where}
\qquad
\delta(\rho)= {\sum}^{*}[\sigma_\alpha]\otimes[\sigma_\beta].
\end{gather*}

In particular when $\omega$ is a~Dunkl connection, we have
\begin{gather}\label{dunkl-curvature-3}
r_\omega(\rho)=-w_\rho {\sum}^{*}h_\alpha h_\beta,
\end{gather}
and the elements $\alpha$ and $\beta$ f\/iguring in the above sum are such that $\alpha\wedge\beta=w_\rho$.
These vectors~$\alpha$ and~$\beta$ all belong to the 2-plane of the rotation $\rho$, and actually the
problem is naturally being reduced to the subgroup of the Coxeter group generated by all the ref\/lections
associated to the above mentioned plane.
It is worth mentioning that Coxeter groups in dimension~2 coincide with the class of dihedral groups.
\end{proof}
\begin{remark}
The fact that curvature forms of connections can only take non-zero values on proper 2-rotations is an
interesting purely quantum phenomenon.
By def\/inition the rotations project to zero by the quantum germs map
$\pi\colon\mathcal{A}\rightarrow\Gamma_{{\rm inv}}$.
So the curvature is not indexed by the internal degrees of freedom of the calculus on the group (as is the
case in classical geometry, where the curvature tensor has values in the Lie algebra of the structure
group) but by objects~$\rho$ `external' to the calculus.
\end{remark}

Let us now consider the very special case of standard Dunkl connections, where
$h_\alpha(x)=\varkappa_\alpha/\langle x,\alpha\rangle$ (see Def\/inition~\ref{definition_standard_dunkl}).
We shall prove the curvature is identically zero in this case as follows from Proposition~1.7(1)
in~\cite{DU}.
For the convenience of the reader, we include an alternative proof, which shares with the original proof
in~\cite{DU} (and other proofs such as in~\cite{CH89}) the reduction to a~fact about the Euclidean plane.
However, the motivation for our proof comes from the consideration of a~geometric property not considered
in~\cite{DU} nor elsewhere, namely the curvature of a~connection.
This key fact is the following interesting property of vectors in the Euclidean plane.
\begin{lemma}[\protect{cf.~\cite[Proposition~1.7]{DU}}] Let us consider a~two-dimensional Euclidean vector space~$\Pi$ with
$\varrho\colon\Pi\rightarrow\Pi$, a~rotation of order $2m$, where $m\geq2$.
$($In other words the rotation $\varrho$ satisfies $\varrho^m=-I$, the central ref\/lection map in $E.)$
Then we have the identity
\begin{gather}
\label{vxrho-sum}
\frac{1}{\langle x,v\rangle\langle x,v\varrho\rangle}+\cdots+\frac{1}{\langle x,v\varrho^{m-2}
\rangle\langle x,v\varrho^{m-1}\rangle}=\frac{1}{\langle x,v\varrho^{m-1}\rangle\langle x,v\rangle},
\end{gather}
where $v\in\Pi$ is a~unit vector and $x\in\Pi$ an arbitrary vector not orthogonal to any of the vectors
$v,v\varrho,\dots,v\varrho^{m-1}$.
\end{lemma}

\begin{proof}
Our proof is based on a~simple recursive sequence of vectors in $\Pi$.
Let $v\in\Pi$ be a~non-zero vector.
Assume that it is associated with a~sequence of vectors $\{v_k\}_{k\ge1}$ in $\Pi$ in such a~way that
\begin{gather}
\label{req-c}
v=c_{k+1}v_k-c_kv_{k+1}
\end{gather}
for each $k\ge1$ and for some sequence of real numbers $\{c_k\}_{k\ge1}$ satisfying $c_1=1$.
Then for each $k\geq2$ the following identity holds when all the denominators are non-zero:
\begin{gather*}
\frac{1}{\langle x,v\rangle\langle x,v_1\rangle}+\frac{1}{\langle x,v_1\rangle\langle x,v_2\rangle}
+\cdots+\frac{1}{\langle x,v_{k-1}\rangle\langle x,v_k\rangle}=\frac{c_k}
{\langle x,v\rangle\langle x,v_k\rangle}.
\end{gather*}
(Actually, this identity also holds for $k=1$ provided that we def\/ine $v_0:=v$.) This identity can easily
be proved inductively by using the above recursive relation~\eqref{req-c}.
We leave that proof to the reader.

Returning to the context of this lemma we def\/ine $v_k=v\varrho^k$ for every integer $k\ge1$.
Then~\eqref{req-c} holds with $c_k=\sin(k\nu)/\sin(\nu)$ and $\nu$ is an angle of a~primitive $m$-th root
of the central ref\/lection map in $\Pi$ (for example $\nu=\pi/m$).
Observing that $c_{m-1}=1$ completes the proof.
\end{proof}
\begin{remark}
For $k=3$ our identity can be derived from the following simple expression involving three arbitrary
vectors $\alpha,\beta,\gamma\in\Pi$, namely 
\begin{gather*}
\frac{1}{\langle x,\alpha\rangle\langle x,\beta\rangle}+\frac{1}
{\langle x,\beta\rangle\langle x,\gamma\rangle}+\frac{1}{\langle x,\gamma\rangle\langle x,\alpha\rangle}
=\frac{\langle x,\alpha+\beta+\gamma\rangle}
{\langle x,\alpha\rangle\langle x,\beta\rangle\langle x,\gamma\rangle},
\end{gather*}
provided that all the denominators are non-zero.
\end{remark}

\begin{theorem}
Suppose that $\omega$ is a~standard Dunkl connection.
Then the curvature tensor $r_\omega$ vanishes identically.
\end{theorem}
\begin{proof}
Applying the formula~\eqref{dunkl-curvature} it is easy to see that the sum~\eqref{dunkl-curvature-3}
reduces to a~linear combination of one or more expressions each of which is a~sum of the type
\begin{gather*}
\left(\frac{1}{\langle x,v\rangle\langle x,v\varrho\rangle}+\cdots+\frac{1}{\langle x,v\varrho^{m-2}
\rangle\langle x,v\varrho^{m-1}\rangle}\right)-\frac{1}{\langle x,v\varrho^{m-1}
\rangle\langle x,v\rangle}.
\end{gather*}
It is important to note that the rotation $\varrho$ is by an angle that is one half the angle of the
rotation $\rho\in G$.
According to~\eqref{vxrho-sum} all these expressions are zero, and thus the curvature $r_\omega$ vanishes
identically.
\end{proof}

A very interesting modif\/ication of standard Dunkl connections arises if we use
\begin{gather*}
h_\alpha(x)/\varkappa_\alpha=\mathrm{coth}
\bigl[\langle x,\alpha\rangle\bigr/2]=\frac{e^{\langle x,\alpha\rangle}+1}{e^{\langle x,\alpha\rangle}-1}.
\end{gather*}
The standard Dunkl connections can be viewed as f\/irst-order approximations of these connections.
The curvature of such connections will not vanish on proper 2-rotations.
However there exists a~particular context where the curvature will have constant values on these rotations.

Let us f\/irst observe that in the above expression the lengths of the root vectors $\alpha$ matter in
a~very essential way (although as we mentioned, in the limit of small vectors $x$ the formula reduces to
the standard Dunkl form).

A particularly interesting context arises when we start from the vectors as in~\eqref{req-c} and put
$c_k=1$ for all $k\ge1$.
In other words we have
\begin{gather*}
v_k-v_{k+1}=v,
\end{gather*}
and so $v_k$ is an arithmetic sequence of vectors, namely $v_{k+1}=v_1-k v$ for all integers $k\ge1$.
We can then apply the following elementary operation: between each two neighbors of the sequence insert
their sum.
\begin{definition}
Such a~sequence will be called a~{\it sequence of arithmetic type}.
\end{definition}

\begin{lemma}
The following identity holds
\begin{gather}
\label{coth-sum}
\sum_{j=0}^k\mathrm{coth}[\langle x,v_j\rangle]\mathrm{coth}[\langle x,v_{j+1}\rangle]=\mathrm{coth}
[\langle x,v\rangle]\mathrm{coth}[\langle x,v_{k+1}\rangle]+k
\end{gather}
for every sequence $v_0=v$, $v_1,\dots,v_k$, $v_{k+1}$ of vectors of arithmetic type. 
\end{lemma}
\begin{proof}
The standard addition formula for the hyperbolic cotangent, namely
\begin{gather*}
\mathrm{coth}(a+c)=\frac{1+\mathrm{coth}(a)\mathrm{coth}(c)}{\mathrm{coth}(a)+\mathrm{coth}(c)},
\end{gather*}
gives
\begin{gather*}
\mathrm{coth}[\langle x,\alpha\rangle]\mathrm{coth}[\langle x,\beta\rangle]+\mathrm{coth}
[\langle x,\beta\rangle]\mathrm{coth}[\langle x,\gamma\rangle]=\mathrm{coth}
[\langle x,\alpha\rangle]\mathrm{coth}[\langle x,\gamma\rangle]+1,
\end{gather*}
where $\beta=\alpha+\gamma$ and we substitute $a=\langle x,\alpha\rangle$, $c=\langle x,\gamma\rangle$ and
$b=\langle x,\beta\rangle=a+c$.
Our main formula now easily follows by applying induction on $k$.
\end{proof}

Our next natural and interesting question is whether sequences of arithmetic type appear in the context of
calculating the curvature of Dunkl connections.
It turns out that this is equivalent to assuming that the root system $R$ is {\it classical} in the sense
of being associated to a~compact semisimple Lie algebra.
\begin{proposition}\quad
\begin{enumerate}\itemsep=0pt
\item[$(i)$] Let us assume that the vectors $\beta+k\alpha$ for $k=0,\dots,m$ $($where $m\ge1$ is an integer$)$ and~$\alpha$ generate a~sequence of arithmetic type which forms a~positive half of a~two-dimensional root
system. 
Then this system is classical and isomorphic to one of the following:
\begin{itemize}\itemsep=0pt
\item The Hexagonal system for $m=1$.
In this case $\alpha$ and $\beta$ have the same length, and the angle between them is $2\pi/3$.

\item The Octagonal system.
This corresponds to $||\beta||=\sqrt{2}||\alpha||$ and $m=2$, or $||\alpha||=\sqrt{2}||\beta||$ and $m=1$,
with the angle between them being $3\pi/4$.

\item The Dodecagonal snowf\/lake system.
In this case we have $||\beta||=\sqrt{3}||\alpha||$ and $m=3$, or $||\alpha||=\sqrt{3}||\beta||$ and $m=1$,
and the angle between $\alpha$ and $\beta$ is $5\pi/6$.
\end{itemize}
In particular, the numbers $1$, $2$ and $3$ are the only possible values for~$m$.

\item[$(ii)$]  If the root system $R$ is such that the above phenomenon occurs for every irreducible
  \mbox{$2$-dimensional} subsystem of~$R$, then~$R$ is classical.
  \end{enumerate}
\end{proposition}

\begin{proof}
Everything follows by applying standard considerations for classifying 2-dimensional classical root
systems~\cite{Serr}.
Let us note in passing that in the case of the hexagonal root system a~set of positive vectors is given by
the initial sequence of the three vectors $\alpha$, $\beta$ and $\beta+\alpha$.
In the case of the dodecagonal system, there always appear additional vectors, and for the octagonal system
the additional vectors appear only when $||\alpha||=\sqrt{2}||\beta||$.
\end{proof}

\begin{remark}
Let us observe that since the system $R$ is classical we do not assume any longer the normalization
condition, according to which all vectors are of the same norm equal to the square root of two.
In fact, for standard Dunkl connections the norm of the vectors is irrelevant, since they appear linearly
in the nominator and the denominator.
In general situations however, it might be more convenient to use some particular normalization.
In those cases, we can write a~`neutral' formula
\begin{gather*}
w_\rho=2\frac{\alpha\wedge\beta}{||\alpha||\,||\beta||}
\end{gather*}
for the volume element of a~proper two-dimensional rotation $\rho$.
Clearly, this generalizes our previous def\/inition.
\end{remark}
\begin{definition}
We say that a~Dunkl connection $\omega=\varpi+\lambda$ has {\it constant curvature} if the values of
$r_\omega$ on proper 2-rotations $\rho$ are of the form
\begin{gather*}
r_\omega(\rho)=-c_\rho w_\rho,
\end{gather*}
where $c_\rho$ are constants.
In other words $c_\rho= {\sum}^{*}h_\alpha(x)h_\beta(x)$ does not depend on $x\in P$.
\end{definition}

\begin{theorem}
If the root system $R$ is classical and a~Dunkl connection $\omega=\varpi+\lambda$ is given by taking
\begin{gather*}
h_\alpha(x)/\varkappa_\alpha=\mathrm{coth}[\langle x,\alpha\rangle/2],
\end{gather*}
then it has constant curvature.
\end{theorem}
\begin{proof}
This follows by a~direct summation and by applying formulas~\eqref{dunkl-curvature} and~\eqref{coth-sum}.
The number of terms in the def\/ining sum for the curvature is half the number of elements in the
corresponding 2-dimensional root subsystem generated by the rotation $\rho$.
\end{proof}

As explained in Appendix~\ref{appendixA},
general covariant derivatives satisfy a~kind of twisted Leibniz rule with the
presence of a~third term involving the non-regularity measure operator $\ell_\omega$.
Let us now calculate explicitly this operator for connections $\omega=\varpi+\lambda$ given by arbitrary
displacements~$\lambda$.
\begin{lemma}
\label{ell-S}
We have
\begin{gather*}
\ell_\omega([s],\varphi)=\lambda[s](\varphi-\varphi_s)
\end{gather*}
for every reflection $s\in G$ and every $\varphi\in\mathfrak{hor}(P)$.
\end{lemma}
\begin{proof}
A direct computation gives
\begin{gather*}
\ell_\omega([s],\varphi)=[s]\varphi-(-)^{\partial\varphi}\varphi^{(0)}\big([s]\circ\varphi^{(1)}\big)
+\lambda[s]\varphi-(-)^{\partial\varphi}\varphi^{(0)}\lambda\big([s]\circ\varphi^{(1)}\big)
=\lambda[s](\varphi-\varphi_s),
\end{gather*}
where we have used the expansion $\varphi^{(0)}\otimes\varphi^{(1)}= \sum\limits_{g\in
G}\varphi_g\otimes g$, the def\/inition of the product in~$\Omega(P)$ and the fact that the right
$\mathcal{A}$-module structure $\circ$ acts on the elements $[s]$ as the scalar multiplication by the
values of functions over $G$ in these ref\/lections, that is
\begin{gather*}
[s]\circ f=f(s)[s]
\end{gather*}
for every $f\in\mathcal{A}$.
Also an expression such as $(-)^{\partial f}$ means that $f$ is a~homogeneous element of a~graded object,
that ${\partial f}$ is its degree and that $(-)\equiv-1$.
\end{proof}

\begin{remark}
In the above lemma, it was not necessary to assume that the calculus is based on ref\/lections.
In fact, the lemma holds for an arbitrary bicovariant and $*$-covariant calculus over~$G$.
In particular we see that the connection based on a~non-zero displacement $\lambda$ is never a~regular connection.
Besides regular connections, there is another important class of connections, called {\it multiplicative connections}.
Their def\/ining property is (see~\cite{D2} and Appendix~\ref{appendixA})
the quadratic identity
\begin{gather*}
\omega\pi\big(r^{(1)}\big)\omega\pi\big(r^{(2)}\big)=0
\end{gather*}
for every $r\in\mathcal{R}$, where $\mathcal{R}\subseteq\ker(\epsilon)$ is the right $\mathcal{A}$-ideal
that determines the calculus~$\Gamma$ over~$G$ (see~\cite{W-diff}). Note that the elements $\pi\big(r^{(1)}\big)\otimes\pi\big(r^{(2)}\big)$ (Sweedler again) provide
precisely the quadratic relations for the universal dif\/ferential envelope $\Gamma^\wedge$ for $\Gamma$
(as discussed in detail in~\cite[Appendix~B]{D1}), and in particular the algebra of left-invariant
dif\/ferential forms $\Gamma_{{\rm inv}}^\wedge$ can be \mbox{obtained} by dividing the tensor algebra
$\Gamma_{{\rm inv}}^\otimes$ by these quadratic relations
(see Appendix~\ref{appendixA}
for more about $\Gamma_{{\rm inv}}$, including its def\/inition). Consequently, the above
quadratic identity means precisely that $\omega$ is extendible to a~multiplicative unital homomorphism
$\omega^\wedge\colon\Gamma^\wedge_{{\rm inv}}\rightarrow\Omega(P)$.
This justif\/ies the terminology `multiplicative connection'.
It is also worthwhile to observe that for a~general connection the expression
$\omega\pi\big(r^{(1)}\big)\omega\pi\big(r^{(2)}\big)$ will always be horizontal for $r\in\mathcal{R}$.
This is yet another purely quantum phenomenon, where we have expressed something horizontal as
a~homogeneous quadratic polynomial involving vertical objects only.
\end{remark}

Now we can relate the multiplicativity of connections with vanishing curvature.
\begin{proposition}
For a~general connection $\omega=\varpi+\lambda$ the following properties are equivalent:
\begin{enumerate}\itemsep=0pt
\item[$(i)$] The connection is multiplicative or in other words there exists a~$($necessarily unique and
$*$-preserving$)$ unital multiplicative extension $\omega^\wedge\colon\Gamma^\wedge_{{\rm
inv}}\rightarrow\Omega(P)$ of $\omega$.

\item[$(ii)$]  The curvature of $\omega$ vanishes on $\mathcal{R}$ or in other words $r_\omega(\rho)=0$ for every
two-dimensional rotation~$\rho$.
\end{enumerate}
\end{proposition}

\begin{proof}
First, it is straightforward to see that the only elements $r\in\mathcal{R}$ for which the expression
$\pi\big(r^{(1)}\big)\otimes\pi\big(r^{(2)}\big)$ can possibly be non-zero are the two-dimensional rotations $r=\rho$.
Second, a~couple of elementary transformations leads us to conclude that
\begin{gather*}
\lambda\pi\big(\rho^{(1)}\big)\varpi\pi\big(\rho^{(2)}\big)+\varpi\pi\big(\rho^{(1)}\big)\lambda\pi\big(\rho^{(2)}\big)
\\
\qquad{}=
\lambda\pi\big(\rho^{(1)}\big)\otimes\pi\big(\rho^{(2)}\big){}-\lambda\pi\big(\rho^{(3)}\big)
\otimes\bigl[\pi\big(\rho^{(1)}\big)\circ\bigl(\kappa\big(\rho^{(2)}\big)\rho^{(4)}\bigr)\bigr]
\\
\qquad{}=
\lambda\pi\big(\rho^{(2)}\big)\otimes\pi\bigl(\kappa\big(\rho^{(1)}\big)\rho^{(3)}\bigr)=0
\end{gather*}
because of the $\mathrm{ad}$-invariance of $\mathcal{R}$ or in other words
$\mathrm{ad}(\mathcal{R})\subseteq\mathcal{R}\otimes\mathcal{A}$.
Together with the obvious identity $\varpi\pi\big(\rho^{(1)}\big)\varpi\pi\big(\rho^{(2)}\big)=0$ this implies that
\begin{gather*}
\omega\pi\big(\rho^{(1)}\big)\omega\pi\big(\rho^{(2)}\big)=\lambda\pi\big(\rho^{(1)}\big)\lambda\pi\big(\rho^{(2)}\big),
\end{gather*}
which is ef\/fectively the third term in the expression for the curvature of the displaced connection.
And that ends the proof.
\end{proof}
\begin{remark}
So when either of the equivalent conditions of the above lemma holds, the curvature behaves more
`classically', namely its internal indices are given by the basis of $\Gamma_{{\rm inv}}$, and so it is
naturally projectable down to $\Gamma_{{\rm inv}}$.
In this case we simply have
\begin{gather*}
r_\omega(s)=D\lambda[s]
\end{gather*}
for every $s\in S$.
In particular, the curvature vanishes if and only if the displacement $\lambda$ is closed.
\end{remark}
\begin{remark}
Let us also observe that the above lemma holds for an arbitrary quantum principal bundle with the
dif\/ferential calculus constructed as a~crossed product of $\mathfrak{hor}(P)$ and $\Gamma_{{\rm
inv}}^\wedge$ in the way described in Appendix~\ref{appendixA}.
\end{remark}

Using the above formula for the non-regularity obstacle $\ell_\omega$, we can f\/ind an explicit expression
for the third term in the quantum Leibniz rule for covariant derivatives.
\begin{theorem}
Let $\lambda$ be an arbitrary displacement map.
Then the associated connection $\omega=\varpi+\lambda$ satisfies
\begin{gather}
\label{lambda-Leibniz}
D_\omega(\varphi\psi)=D_\omega(\varphi)\psi+(-)^{\partial\varphi}\varphi D_\omega(\psi)-\sum_{s\in S}
\lambda[s](\varphi-\varphi_s)(\psi-\psi_s)
\end{gather}
for all $\varphi,\psi\in\mathfrak{hor}(P)$. 

\end{theorem}
\begin{proof}
A direct computation, using~\eqref{Fw-G}, gives
\begin{gather*}
D_\omega(\varphi\psi)=D_\omega(\varphi)\psi+(-)^{\partial\varphi}
\varphi D_\omega(\psi)+(-)^{\partial\varphi}\varphi^{(0)}\ell_\omega\bigl(\pi\big(\varphi^{(1)}\big),\psi\bigr)
\\
\hphantom{D_\omega(\varphi\psi)}{}=D_\omega(\varphi)\psi+(-)^{\partial\varphi}\varphi D_\omega(\psi)+(-)^{\partial\varphi}
\varphi\ell_\omega\bigl([\epsilon],\psi)+(-)^{\partial\varphi}\sum_{s\in S}
\varphi_s\ell_\omega\bigl([s],\psi\bigr)
\\
\hphantom{D_\omega(\varphi\psi)}{}=D_\omega(\varphi)\psi+(-)^{\partial\varphi}\varphi D_\omega(\psi)+(-)^{\partial\varphi}\sum_{s\in S}
(\varphi_s-\varphi)\ell_\omega\bigl([s],\psi\bigr)
\\
\hphantom{D_\omega(\varphi\psi)}{}=D_\omega(\varphi)\psi+(-)^{\partial\varphi}\varphi D_\omega(\psi)+(-)^{\partial\varphi}\sum_{s\in S}
(\varphi_s-\varphi)\lambda[s](\psi-\psi_s).
\end{gather*}
Besides elementary transformations, we have applied Lemma~\ref{ell-S} and the general Leibniz rule for
covariant derivatives.
The notation $(-)^{\partial\varphi}$ was explained above.
\end{proof}

\begin{remark}
This result generalizes to horizontal forms Theorem~\ref{Leibniz-F}, which is only for functions.
\end{remark}

We now apply these results regarding arbitrary displacements to Dunkl connections, where the displacements
are of the form~\eqref{generic-lambda}.
In this special case the classical Leibniz rule is fulf\/illed for a~large class of horizontal forms in
conjunction with arbitrary forms.
As we can see explicitly from~\eqref{lambda-Leibniz} if one of the forms~$\varphi$ or~$\psi$ is
$G$-invariant, in other words if it belongs to the forms on the base space~$M$, then the classical Leibniz
rule holds.
\begin{proposition}
Let $\omega=\varpi+\lambda$ be a~Dunkl connection.
\begin{enumerate}\itemsep=0pt
\item[$(i)$] If $\theta$ is a~coordinate one-form on $P$ then
\begin{gather}
\label{D-theta}
D_\omega(\theta)=0
\end{gather}
and more generally we have
\begin{gather}
\label{D-theta-linear}
D_\omega(\varphi\theta)=D_\omega(\varphi)\theta
\end{gather}
for each $\varphi\in\mathfrak{hor}(P)$.

\item[$(ii)$] Let $\mathcal{W}$ be the subalgebra of $\mathfrak{hor}(P)$ generated by the elements of
$\Omega(M)$ and the coordinate one-forms $\theta$.
Then $\mathcal{W}$ is $*$-invariant, $D$-invariant and $G$-invariant in the sense that
$F_\wedge(\mathcal{W})\subseteq\mathcal{W}\otimes\mathcal{A}$.
The action of the operator $D_\omega$ coincides with the action of $D$ on $\mathcal{W}$.
We have
\begin{gather}
D_\omega(\varphi\psi)=D_\omega(\varphi)\psi+(-)^{\partial\varphi}\varphi D(\psi),
\nonumber\\
D_\omega(\psi\varphi)=D(\psi)\varphi+(-)^{\partial\psi}\psi D_\omega(\varphi)\label{W-Leibniz}
\end{gather}
for each $\varphi\in\mathfrak{hor}(P)$ and $\psi\in\mathcal{W}$.
\end{enumerate}
\end{proposition}

\begin{proof}
By def\/inition, being dif\/ferentials of coordinates, the coordinate one-forms $\theta$ are closed; in
other words $D(\theta)=0$.
Now, the displacement $\lambda[s]$ for a~Dunkl connection is proportional to the coordinate one-form
$\alpha$, where $s=\sigma_\alpha$ and the dif\/ference $\theta-\theta_s$ is also proportional to $\alpha$.
This means that the displacement part of the covariant derivative vanishes too, and hence~\eqref{D-theta}
holds.
Similarly, applying~\eqref{lambda-Leibniz} we conclude that~\eqref{D-theta-linear} holds.
As a~direct consequence of this property, we conclude that $D$ and $D_\omega$ coincide on $\mathcal{W}$.

Finally, the properties in~\eqref{W-Leibniz} follow from observing that they hold for $\psi$ either in~$\Omega(M)$ or equal to a~coordinate form (and for arbitrary horizontal form~$\varphi$).
Then we use the fact that all such horizontal forms $\psi$ satisfying the classical Leibniz rule for
arbitrary $\varphi\in\mathfrak{hor}(P)$ form a~subalgebra of $\mathfrak{hor}(P)$.
\end{proof}

\begin{remark}
It is worth mentioning that, as a~direct corollary of the above proposition, we have
\begin{gather*}
D_\omega(\varphi\Theta)=D_\omega(\varphi)\Theta
\end{gather*}
for every $\Theta$ belonging to the $2^n$-dimensional subalgebra generated by the coordinate one-forms.
\end{remark}

We shall now analyze the coordinate representation of the covariant derivative maps and their relation with
the quantum curvature tensor.
In particular, we shall see that covariant directional derivatives commute among themselves when the
curvature of a~Dunkl connection vanishes.

Let us consider the canonical, Euclidean coordinates $x_1,\dots,x_n$ as real-valued functions on~$P$.
For an arbitrary connection $\omega$ we have a~natural decomposition
\begin{gather}
\label{D-b-coord}
D_\omega(b)=\sum_{k=1}^n\partial_\omega^k(b)\theta_k,
\qquad
\theta_k=D(x_k),
\end{gather}
where $b\in\mathcal{B}$.
The linear maps $\partial_\omega^k\colon\mathcal{B}\rightarrow\mathcal{B}$, known as {\it covariant partial
derivatives}, completely determine the operator $D_\omega$.
This is because a~horizontal $m$-form $\varphi$ can be decomposed naturally as
\begin{gather*}
\varphi=\sum_{i_1<\cdots<i_m}b_{i_1\dots i_m}\theta_{i_1}\cdots\theta_{i_m}
\end{gather*}
with $b_{i_1\dots i_m}\in\mathcal{B}$ and because $D_\omega$ is right linear over the subalgebra generated
by the coordinate forms.
Explicitly, for Dunkl connections these covariant partial derivatives are given for $b\in\mathcal{B}$,
$x\in P$ and $k=1,\dots,n$ by
\begin{gather}
\label{dkbx}
\partial_\omega^k(b)(x)=\partial^k(b)(x)+i\sum_{\alpha\in{R}^+}
\psi_\alpha\bigl[\langle x,\alpha\rangle\bigr]\bigl(b(x)-b(x\sigma_\alpha)\bigr)\alpha_k,
\end{gather}
where $\alpha=(\alpha_1,\dots,\alpha_k,\dots,\alpha_n)\in{R}^+$ and where $\partial^k=\partial/\partial
x_k$ is the usual partial derivative.
\begin{remark}
The title of this paper refers to equation~\eqref{dkbx} when we take $\psi_\alpha$ to be as
in~\eqref{standard_dunkl}.
\end{remark}

The coordinate representation of the curvature tensor is given by
\begin{gather*}
r_\omega(\rho)=\frac{1}{2}\sum_{k,l=1}^nr_\omega^{kl}(\rho)\theta_k\theta_l,
\end{gather*}
where $r_\omega^{kl}(\rho)=-r_\omega^{lk}(\rho)$ are smooth functions on $P$ and $\rho\in G$ an arbitrary
proper two-dimensional rotation.
\begin{proposition}
Let $\omega=\varpi+\lambda$ be a~Dunkl connection.
\begin{enumerate}\itemsep=0pt
\item[$(i)$] We have
\begin{gather}
\label{r-comm}
[\partial_\omega^k,\partial_\omega^l](b)+\sum_\rho b_\rho r^{kl}_\omega(\rho)=0
\end{gather}
for each $b\in\mathcal{B}$.
Here the sum is taken over all proper $2$-rotations $\rho$ of the Coxeter group~$G$.

\item[$(ii)$] In particular, the curvature of the connection $\omega$ vanishes if and only if
\begin{gather*}
\partial_\omega^k\partial_\omega^l=\partial^l_\omega\partial_\omega^k
\end{gather*}
for every $k,l\in\{1,\dots,n\}$.
\end{enumerate}
\end{proposition}
\begin{proof}
We recall a~general expression connecting the curvature and the square of the covariant derivative:
\begin{gather*}
D_\omega^2(\varphi)=-\varphi^{(0)}r_\omega\big(\varphi^{(1)}\big)=-\sum_\rho\varphi_\rho r_\omega(\rho),
\end{gather*}
where the sum on the right hand side runs over all of the proper two-dimensional rotations $\rho$, these
being the only elements from $G$ on which the curvature tensor might possibly be non-zero.
Using~\eqref{D-b-coord} and~\eqref{D-theta-linear} as well as applying the above result about the curvature
tensor for $b\in\mathcal{B}$, we obtain
\begin{gather*}
D_\omega^2(b)=-\sum_\rho b_\rho r_\omega(\rho)=-\frac{1}{2}\sum_\rho\!\sum_{k,l=1}^nb_\rho r_\omega^{kl}
(\rho)\theta_k\theta_l=\sum_{k=1}^n D_\omega\bigl(\partial_\omega^k(b)\theta_k\bigr)
\\
\hphantom{D_\omega^2(b)}{}
=\sum_{k=1}^n D_\omega\bigl(\partial_\omega^k(b)\bigr)\theta_k=\frac{1}{2}\sum_{k,l=1}
^n[\partial_\omega^k,\partial_\omega^l](b)\theta_k\theta_l.
\end{gather*}
Hence~\eqref{r-comm} holds.
In the special case when the curvature vanishes, we conclude that the partial covariant derivatives commute.
Conversely, if the partial covariant derivatives commute, then the square of the covariant derivative
vanishes on $\mathcal{B}$ and thus it vanishes completely.
This means that the curvature is zero.
\end{proof}

\begin{remark}
It is worth recalling that in establishing the above properties, we have used in an essential way the fact
that $\mathfrak{hor}(P)$ is the classical graded-commutative algebra of dif\/ferential forms, and that the
quantum connection $\omega$ is given by a~very special kind of displacement map~$\lambda$ for which
$\lambda[\sigma_\alpha]$ is always proportional to~$\alpha$.
On the other hand, in the most general context of arbitrary quantum principal $G$-bundles the curvature~$r_\omega$ may only have `purely quantum' non-zero components on the neutral element where
\begin{gather*}
r_\omega(\epsilon)=\lambda[\epsilon]^2+\sum_{s\in S}\lambda[s]^2
\end{gather*}
and on two-dimensional rotations, including the non-proper ones of order two, besides the `standard'
components of $r_\omega$ on the elements $s\in S$ projecting down to the canonical basis in the space
$\Gamma_{{\rm inv}}$.
On the 2-rotation elements we would have our already familiar expression
\begin{gather*}
r_\omega(\rho)= {\sum}^{*}\lambda[\sigma_\alpha]\lambda[\sigma_\beta].
\end{gather*}
All of this is a~consequence of the nature of the choice of the dif\/ferential calculus on $G$.
So we can consider it to be an interesting, purely quantum phenomenon that in the context of our main
interest (where the bundle space is classical but the calculus is quantum and $\omega$ is a~Dunkl
connection) that the curvature can only assume non-zero values on proper rotations.
\end{remark}

\section{Concluding observations}

The upshot of this article has been to view Dunkl operators in a~much wider context than in previous
research.
That general context, quantum principal bundles with a~given connection, is quite broad and we would like
to draw attention to the particular niche that Dunkl operators occupy in it.
Some basic properties when viewed this way are the following:
\begin{itemize}\itemsep=0pt

\item The spaces of the bundle are classical.

\item The structural group of the bundle is classical, namely a~f\/inite Coxeter group, even though the
theory allows in general any quantum group, f\/inite or not.

\item The dif\/ferential calculi in the f\/iber space and in the total space are not classical, though the
dif\/ferential calculus in the base space is classical.

\item The covariant derivatives form a~commutative family of operators.

\end{itemize}

Moreover, the connection is a~perturbation by a~natural displacement term of the classical Levi-Civita
connection of curvature zero in an open subset of a~Euclidean space.
This perturbed connection again has curvature zero.
Most importantly, this connection is realized as a~specif\/ic case of connections in non-commutative
geometry, these having been def\/ined and studied earlier and independently of this work.
Others have noted that Def\/inition~\ref{define-dunkl-operator} and other similar def\/initions bear
a~passing resemblance to classical covariant derivatives and so have simply dubbed such operators to be
`connections' though they are are not classical due to the non-local term.
We consider our approach to be more fundamental.

We have used `classical' here to refer to the context of classical dif\/ferential geometry.
The theory of quantum principal bundles allows the possibility of changes in all of these properties.
As one example let us note that any f\/inite group gives rise to a~quantum group in the sense of Woronowicz.
So we foresee many avenues for future research, both in geometry and analysis, based on this point of view.

In this paper we have assumed that the quantum connections $\omega\colon\Gamma_{{\rm
inv}}\rightarrow\Omega(P)$ are real in the sense of intertwining the $*$-structures on $\Gamma_{{\rm inv}}$
and $\Omega(P)$.
This condition corresponds to the standard reality property of the connection form in classical
dif\/ferential geometry.
Because of it, we have the appearance of the imaginary unit $i=\sqrt{-1}$ in several expressions in
contrast with the usual formulas in Dunkl theory.
From the algebraic viewpoint this reality condition is not essential and the rest of our formalism works
without it as well.
Geometrically, this corresponds to playing with the elements of the complexif\/ied af\/f\/ine space of
connections.
Every such complex connection can be uniquely decomposed as $\omega_C=\omega+i\lambda$ where $\omega$ and
$\lambda$ are a~standard (real) connection and a~displacement map, respectively.

Our main example of a~principal bundle $P$ is not very interesting topologically, since it is a~trivial
bundle over the base $M$, which itself is the Weyl chamber associated to the root system~$R$.
Much more interesting topological situations~\cite{DUO} arise when we consider a~complex Euclidean space
(the complexif\/ication of the initial Euclidean space $E$).

Several generalizations of the original Dunkl theory have appeared over the years in the literature.
We expect that these also will f\/it into the general, basic framework presented in this paper.
Among them we mention the Dunkl--Cherednik operators~\cite{opdam}, the Jacobi--Dunkl operators~\cite{CGM},
the generalized Dunkl operators in~\cite{BKO} and~\cite{DOSS} and the complex Dunkl operators~\cite{DUO}.
The last reference concerns a~situation where the manifolds have complex structure.
In such a~context we could deal with complex scalar products explicitly, and so with complex connections
and displacements.
Moreover, our main results and formulas would still remain valid in this context.

Our methods reconf\/irm the fundamental signif\/icance of the general concept of curvature for the geometry
of quantum spaces~-- including the case of classical spaces equipped with a~quantum dif\/ferential calculus.
An interesting possible `complementary' application of the formalism would be to study algebraic structures
whose properties are interpreted as a~manifestation of the non-triviality of the curvature of a~quantum
connection.
For example, the groupoid relativity theory of~\cite{zbyszek} provides a~natural framework for such
developments, in which the non-associativity of the relative velocities can be linked with the hyperbolic
curvature (at both classical and quantum levels).

Another interesting topic for further research would be to investigate arbitrary classical and quantum
principal bundles whose structure group is a~Coxeter group.
This group can be equipped with the same dif\/ferential calculus based on ref\/lections as we have done in
this paper or with any other dif\/ferent calculus as outlined in the general theory in the appendices.

\appendix

\section{Quantum principal bundles}\label{appendixA}

In this appendix we review general properties of quantum principal bundles with the emphasis on the
formalism of connections and covariant derivatives.
More detailed explanations and proofs can be found in~\cite{D2,D3}.

We f\/irst recall the def\/inition and interpretation for a~quantum principal bundle
$P=(\mathcal{B},\iota,F)$, which is a~$*$-algebra $\mathcal{B}$, an inclusion $\iota$ and a~co-action $F$.
Here $\mathcal{B}$ def\/ines $P$ as a~quantum space, the total space of the bundle.
There is a~symmetry of $P$ by a~quantum group $\mathcal{A}$ (also denoted as $G$ since it corresponds to
the Lie group of a~classical principal bundle) via a~linear map
$F\colon\mathcal{B}\rightarrow\mathcal{B}\otimes\mathcal{A}$.
Strictly speaking, the dual concept of an action is a~{\it co-action}, and so this is how one should refer
to the map $F$.
However, we will often speak of $F$ and similar maps as actions, since classically they correspond by
duality to honest-to-goodness actions.
For our present purposes a~quantum group is a~f\/inite-dimensional Hopf algebra with unit.
Also the base space algebra $\mathcal{V}$, corresponding to a~quantum space $M$, is def\/ined as all the
$F$-invariant elements of $\mathcal{B}$, namely, $\mathcal{V}\equiv\{b\in\mathcal{B}\,|\,F(b)=b\otimes1\}.
$ The inclusion map $\mathcal{V}\hookrightarrow\mathcal{B}$, denoted~$\iota$ above, is interpreted as
a~`f\/ibration' of~$P$ over~$M$.

The dif\/ferential calculus on $P$ is specif\/ied by a~graded dif\/ferential $*$-algebra $\Omega(P)$ over
$\mathcal{B}$.
Let us denote by $d_{\!P}\colon\Omega(P)\rightarrow\Omega(P)$ the corresponding dif\/ferential.
We assume that there is a~dif\/ferential morphism
$\widehat{F}\colon\Omega(P)\rightarrow\Omega(P)\mathbin{\widehat{\otimes}}\Gamma^\wedge$, itself an action,
extending the action $F$.
Here~$\Gamma^\wedge$ is the enveloping dif\/ferential algebra of a~given bicovariant $*$-covariant
f\/irst-order calculus~$\Gamma$ over~$G$, as explained and constructed in~\cite[Appendix~B]{D1}.
We have
explicitly these def\/ining properties of an action:
\begin{gather*}
(\mathrm{id}\otimes\epsilon)\widehat{F}=\mathrm{id},
\qquad
\big(\widehat{F}\otimes\mathrm{id}\big)\widehat{F}=\big(\mathrm{id}\otimes\widehat{\phi}\,\big)\widehat{F},
\end{gather*}
where $\widehat{\phi}\colon\Gamma^\wedge\rightarrow\Gamma^\wedge\mathbin{\widehat{\otimes}}\Gamma^\wedge$
is the graded dif\/ferential extension of the coproduct $\phi$ of $\mathcal{A}$ and the counit $\epsilon$
is extended trivially.

Alternatively, we can use the higher-order calculus of~\cite{W-diff} based on the braided exterior algebra
associated to $\Gamma$.
These two higher order calculi are the maximal (the former) and the minimal (the latter) objects in the
category of all higher-order calculi extending $\Gamma$ for which the coproduct map extends to the
graded-dif\/ferential level (necessarily uniquely).
It is also worth mentioning that such a~property implies the bicovariance of $\Gamma$.

The intuitive geometrical interpretation of $\widehat{F}$ and $\widehat{\phi}$ is that they are pullbacks
at the level of dif\/ferential forms of the right action $P\times G\rightsquigarrow P$ and the
multiplication $G\times G\rightsquigarrow G$.

If a~dif\/ferential calculus $\Omega(P)$ satisfying the mentioned properties is given, then the algebra of
horizontal forms $\mathfrak{hor}(P)$ can be def\/ined as
\begin{gather*}
\mathfrak{hor}(P)=\bigl\{w\in\Omega(P) \big \vert \widehat{F}(w)\in\Omega(P)\otimes\mathcal{A}\bigr\},
\end{gather*}
and the restriction of $\widehat{F}$ to $\mathfrak{hor}(P)$ induces a~map
$F_\wedge\colon\mathfrak{hor}(P)\rightarrow\mathfrak{hor}(P)\otimes\mathcal{A}$.
The interpretation is that the horizontal forms exhibit trivial dif\/ferential properties along the
vertical f\/ibers of the bundle.
In the spirit of this interpretation the calculus on the base quantum space $M$ is given by
$F_\wedge$-invariant elements of $\mathfrak{hor}(P)$ or, to put it in equivalent terms, by the
$\widehat{F}$-invariant subalgebra $\Omega(M)$ of $\Omega(P)$.
It is a~dif\/ferential $*$-subalgebra of the full calculus.
We shall write $d_{\!M}\colon\Omega(M)\rightarrow\Omega(M)$ for the corresponding restricted dif\/ferential.

It is worth observing that we have $\Omega^0(P)=\mathcal{B}=\mathfrak{hor}^0(P)$ and
$\Omega^0(M)=\mathcal{V}$.
As a~dif\/ferential algebra we assume that $\mathcal{B}$ generates $\Omega(P)$.
In other words, the $n$-th grade forms are expressible as
\begin{gather*}
\Omega^n(P)=\Bigl\{\sum bd_{\!P}(b_1)\cdots d_{\!P}(b_n)\Bigr\}=\Bigl\{\sum d_{\!P}(b_1)\cdots d_{\!P}
(b_n)b\Bigr\}.
\end{gather*}
Somewhat surprisingly $\Omega(M)$ will not in general be generated by $\mathcal{V}$, although in some very
important special contexts this property will be fulf\/illed.

Let us now consider a~connection $\omega\colon\Gamma_{{\rm inv}}\rightarrow\Omega^1(P)$ on $P$, where
$\Gamma_{{\rm inv}}$ is by def\/inition all the left-invariant elements in the above f\/irst order
dif\/ferential calculus~$\Gamma$
(see~\cite{D2}). By def\/inition~$\omega$ is a~linear map such that for all $\vartheta\in\Gamma_{{\rm
inv}}$ we have
\begin{gather*}
\omega\big(\vartheta^{*}\big)=\omega(\vartheta)^{*},
\qquad
\widehat{F}\omega(\vartheta)=(\omega\otimes\mathrm{id})\mathrm{ad}(\vartheta)+1\otimes\vartheta.
\end{gather*}
The above conditions correspond to the classical idea~\cite{KN} of a~$\mathrm{lie}(G)$-valued
pseudotensorial real one-form which maps fundamental vertical vector f\/ields back into their generators.
The adjoint action $\mathrm{ad}\colon\Gamma_{{\rm inv}}\rightarrow\Gamma_{{\rm inv}}\otimes\mathcal{A}$ of
$G$ on $\Gamma_{{\rm inv}}$ is given by
\begin{gather*}
\mathrm{ad}\,\pi(a)=\pi\big(a^{(2)}\big)\otimes\kappa(a^{(1)})a^{(3)}
\end{gather*}
and $\pi\colon\mathcal{A}\rightarrow\Gamma_{{\rm inv}}$ is the corresponding `quantum germs' projection map
$\pi(a)=\kappa\big(a^{(1)}\big)d\big(a^{(2)}\big)$, using Sweedler's notation twice.
We have
\begin{gather*}
\pi(ab)=\epsilon(a)\pi(b)+\pi(a)\circ b,
\\
\pi(a)^{*}=-\pi[\kappa(a)^{*}],
\\
d\pi(a)=-\pi\big(a^{(1)}\big)\pi\big(a^{(2)}\big),
\\
d(a)=a^{(1)}\pi\big(a^{(2)}\big),
\\
d\kappa(a)=-\pi\big(a^{(1)}\big)\kappa\big(a^{(2)}\big),
\end{gather*}
where $\circ$ is the canonical right $\mathcal{A}$-module structure on $\Gamma_{{\rm inv}}$.
For $\vartheta\in\Gamma_{{\rm inv}}$ and $a~\in\mathcal{A}$ it is given by
\begin{gather*}
\vartheta\circ a=\kappa\big(a^{(1)}\big)\,\vartheta a^{(2)}.
\end{gather*}
We also have $\pi(a)\circ b=\pi(ab)-\epsilon(a)\pi(b)$ for all $a,b\in\mathcal{A}$. 

The covariant derivative of $\omega$ is the map
$D_\omega\colon\mathfrak{hor}(P)\rightarrow\mathfrak{hor}(P)$ def\/ined by
\begin{gather}
\label{def-D}
D_\omega(\varphi)=d_{\!P}(\varphi)-(-)^{\partial\varphi}\varphi^{(0)}\omega\pi\big(\varphi^{(1)}\big).
\end{gather}
(The notation $\partial\varphi$ implies that the formula is valid for homogeneous elements $\varphi$ of
degree $\partial\varphi$.
Accordingly, $(-)^{\partial\varphi}$ is either $+1$ or $-1$.) Then~\eqref{def-D} is a~kind of covariant
perturbation of $d_{\!P}$ by `vertical' terms, so that the resulting expression becomes horizontal.
In the same spirit we also def\/ine a~`mirror' derivative by moving the germs part to the left of $\varphi$
and keeping everything horizontal, namely
\begin{gather*}
D_\omega'(\varphi)=d_{\!P}(\varphi)+\bigl\{\omega\pi\kappa^{-1}\big(\varphi^{(1)}\big)\bigr\}\varphi^{(0)}.
\end{gather*}
These two derivatives are related by the expression
\begin{gather*}
D_\omega'(\varphi)=D_\omega(\varphi)+\ell_\omega\bigl(\pi\kappa^{-1}\big(\varphi^{(1)}\big),\varphi^{(0)}\bigr),
\end{gather*}
where $\ell_\omega\colon\Gamma_{{\rm inv}}\times\mathfrak{hor}(P)\rightarrow\mathfrak{hor}(P)$ is the {\it
regularity deviation} measure~\cite{D-qclass}.
This is a~kind of twisted commutator between connections and horizontal forms given by
\begin{gather*}
\ell_\omega(\vartheta,\varphi)=\omega(\vartheta)\varphi-(-)^{\partial\varphi}\varphi^{(0)}
\omega\big(\vartheta\circ\varphi^{(1)}\big).
\end{gather*}
\begin{remark}
As explained in~\cite{D2} and in much more detail in~\cite{D3} and~\cite{D-qclass}, there are very special
connections, called {\it regular connections}, characterized by the equation $\ell_\omega=0$, in other
words, a~kind of twisted commutation relation between a~connection form and its horizontal forms.
If a~bundle admits a~regular connection, then the theory of characteristic classes \mbox{assumes} \mbox{a~particularly}
simple form that is a~braided version of classical Weil theory.
In general, a~connection will be not regular, and the operator $\ell_\omega$ in a~sense measures the
deviation from regularity.
It is interesting to observe that the values of $\ell_\omega$ are always horizontal, in spite of the fact
that it contains the vertical object $\omega$.
This is a~purely quantum phenomenon.
In classical geometry horizontal forms can not include in a~non-trivial algebraic way the components of
vertical objects such as connection forms.
\end{remark}

Then the following identities hold:
\begin{gather*}
\ell_\omega(\vartheta,\varphi\psi)=\ell_\omega(\vartheta,\varphi)\psi+(-)^{\partial\varphi}\varphi^{(0)}
\ell_\omega\big(\vartheta\circ\varphi^{(1)},\psi\big),
\\
-\ell_\omega(\vartheta,\varphi)^{*}=\ell_\omega\big(\vartheta^{*}\circ\kappa\big(\varphi^{(1)}\big)^{*},\varphi^{(0)*}\big),
\\
F^\wedge\ell_\omega(\vartheta,\varphi)=\ell_\omega\big(\vartheta^{(0)},\varphi^{(0)}\big)\otimes\vartheta^{(1)}
\varphi^{(1)}.
\end{gather*}
By construction the operators $D_\omega$ and $D_\omega'$ both extend the dif\/ferential
\begin{gather*}
d_{\!M}\colon \ \Omega(M)\rightarrow\Omega(M).
\end{gather*}
We have that
\begin{gather*}
D_\omega(\varphi\psi)=D_\omega(\varphi)\psi+(-)^{\partial\varphi}
\varphi D_\omega(\psi)+(-)^{\partial\varphi}\varphi^{(0)}\ell_\omega\bigl\{\pi\big(\varphi^{(1)}\big),\psi\bigr\},
\\
D_\omega'(\varphi\psi)=D_\omega'(\varphi)\psi+(-)^{\partial\varphi}
\varphi D_\omega'(\psi)+\ell_\omega\bigl\{\pi\kappa^{-1}\big(\psi^{(1)}\big){\circ}\kappa^{-1}\big(\varphi^{(1)}
\big),\varphi^{(0)}\bigr\}\psi^{(0)}.
\end{gather*}
These generalize the standard Leibniz rules.
The restricted Leibniz rules
\begin{gather*}
D_\omega(w\varphi)=d_{\!M}(w)\varphi+(-)^{\partial w}wD_\omega(\varphi),
\\
D'_\omega(\varphi w)=D_\omega'(\varphi)w+(-)^{\partial\varphi}\varphi d_{\!M}(w)
\end{gather*}
hold for $w\in\Omega(M)$ and $\varphi\in\mathfrak{hor}(P)$.
These two derivatives are mutually conjugate in the sense that $*D_\omega*=D_\omega'$.

The operator $\ell_\omega$ is completely determined by the covariant derivative map $D_\omega$ since we have
\begin{gather*}
\ell_\omega(\pi(a),\varphi)=[a]_1D_\omega\bigl\{[a]_2\varphi\bigr\}-[a]_1D_\omega\bigl\{[a]_2\bigr\}\varphi,
\end{gather*}
where $\varphi\in\mathfrak{hor}(P)$ and $a\in\ker(\epsilon)$.
Also we used the symbol $[a]_1\otimes[a]_2$ to denote the value of the translation map
$\tau\colon\mathcal{A}\rightarrow\mathcal{B}\otimes_{\mathcal{V}}\mathcal{B}$ on $a\in\mathcal{A}$
(see~\cite{D3} for more details). It is also worth mentioning that the regularity property of $\omega$
(def\/ined as the vanishing condition $\ell_\omega=0$) is equivalent to the standard Leibniz rules for
$D_\omega$ or $D_\omega^{*}$.
If this is the case, then the covariant derivative is also hermitian, that is, $*D_\omega=D_\omega*$.

Another fundamental object naturally associated to $\omega$ is its {\it curvature}.
It can be expressed in terms of the square of the covariant derivative operator.
The restricted Leibniz rules over $\Omega(M)$ imply that the square of the covariant derivative will always
be a~left and right $\Omega(M)$-linear map.
More precisely, we have
\begin{gather*}
D_\omega^2(\varphi)=-\varphi^{(0)}r_\omega\big(\varphi^{(1)}\big),
\qquad
r_\omega(a)=-[a]_1D^2_\omega[a]_2
\end{gather*}
with the `curvature tensor' $r_\omega\colon\mathcal{A}\rightarrow\mathfrak{hor}(P)$ given by a~quantum
version of the classical structure equation
\begin{gather}
\label{romega-structure}
r_\omega(a)=d_{\!P}\omega\pi(a)+\omega\pi\big(a^{(1)}\big)\omega\pi\big(a^{(2)}\big).
\end{gather}
It is instructive to compute directly the square of the covariant derivative.
We have
\begin{gather*}
D_\omega^2(\varphi)=d_{\!P}D_\omega(\varphi)+(-)^{\partial\varphi}D_\omega\big(\varphi^{(0)}
\big)\omega\pi\big(\varphi^{(1)}\big)
\\
\hphantom{D_\omega^2(\varphi)}{}
=d_{\!P}\bigl(d_{\!P}(\varphi)-(-)^{\partial\varphi}\varphi^{(0)}\omega\pi\big(\varphi^{(1)}\big)\bigr)
+(-)^{\partial\varphi}d_{\!P}\big(\varphi^{(0)}\big)\omega\pi\big(\varphi^{(1)}\big)\\
\hphantom{D_\omega^2(\varphi)=}{}
-\varphi^{(0)}\omega\pi\big(\varphi^{(1)}
\big)\omega\pi\big(\varphi^{(2)}\big)
\\
\hphantom{D_\omega^2(\varphi)}{}
=-\varphi^{(0)}d_{\!P}\omega\pi\big(\varphi^{(1)}\big) -\varphi^{(0)}\omega\pi\big(\varphi^{(1)}
\big)\omega\pi\big(\varphi^{(2)}\big)=-\varphi^{(0)}r_\omega\big(\varphi^{(1)}\big)
\end{gather*}
with $r_\omega$ given by the above structure equation.
Furthermore, the following identities hold:
\begin{gather*}
D_\omega r_\omega(a)+\ell_\omega\bigl\{\pi\big(a^{(1)}\big),r_\omega\big(a^{(2)}\big)\bigr\}=0,
\\
r_\omega(a)^{*}=-r_\omega[\kappa(a)^{*}],
\\
F^\wedge r_\omega(a)=r_\omega\big(a^{(2)}\big)\otimes\kappa\big(a^{(1)}\big)a^{(3)}.
\end{gather*}
These generalize the corresponding formulas from classical dif\/ferential geometry.
In particular, the f\/irst of these identities is a~quantum version of the classical Bianchi identity.
The second identity is the reality property for the curvature, and the third establishes its transformation
rule, which classically corresponds to the tensoriality property of the curvature tensor.

It is worth observing that in general the map $r_\omega$ will not be projectable down to quantum germs in
contrast with the classical situation.
So this is a~purely quantum phenomenon which can be interpreted as a~kind of `inadequacy' of the calculus
$\Gamma$ for the bundle $P$.

We also have this rule for covariantly dif\/ferentiating the regularity obstacle:
\begin{gather*}
D_\omega\ell_\omega(\vartheta,\varphi)=r_\omega(a)\varphi-\varphi^{(0)}r_\omega\big(a\varphi^{(1)}
\big)-\ell_\omega(\vartheta,D_\omega(\varphi)){}-\ell_\omega\bigl[\pi\big(a^{(1)}\big),\ell_\omega\bigl(\pi\big(a^{(2)}
\big),\varphi\bigr)\bigr].
\end{gather*}
In the above formula $\vartheta=\pi(a)$ and $a\in\ker(\epsilon)$.

Let us now consider a~linear map $\lambda\colon\Gamma_{{\rm inv}}\rightarrow\mathfrak{hor}^1(P)$ satisfying
\begin{gather*}
\lambda\big(\vartheta^{*}\big)=\lambda(\vartheta)^{*}
\qquad
\mathrm{and}
\qquad
F_\wedge\lambda(\vartheta)=(\lambda\otimes\mathrm{id})\mathrm{ad}(\vartheta)
\end{gather*}
for each $\vartheta\in\Gamma_{{\rm inv}}$.
Such maps are can be interpreted as connection displacements.
They form the vector space (in general, inf\/inite-dimensional) associated to the af\/f\/ine space of all
connections.

So it is interesting to ask what is the relation between the covariant derivatives and curvature tensors of
two connections $\omega$ and $\omega+\lambda$.
We used these formulas in the main text to establish the principal properties of Dunkl operators.
We have
\begin{gather*}
r_{\omega+\lambda}(a)=r_\omega(a)+D_\omega\lambda\pi(a)+\lambda\pi\big(a^{(1)}\big)\lambda\pi\big(a^{(2)}
\big)+\ell_\omega\bigl(\pi\big(a^{(1)}\big),\lambda\pi\big(a^{(2)}\big)\bigr).
\end{gather*}

Indeed, applying the structure equation~\eqref{romega-structure} for the displaced connection
$\omega+\lambda$, and then performing some further elementary transformations, we obtain
\begin{gather*}
r_{\omega+\lambda}(a)=d_{\!P}\omega\pi(a)+d_{\!P}\lambda\pi(a)+\omega\pi\big(a^{(1)}\big)\omega\pi\big(a^{(2)}\big)
\\
\hphantom{r_{\omega+\lambda}(a)=}{}
+\lambda\pi\big(a^{(1)}\big)\omega\pi\big(a^{(2)}\big) +\omega\pi\big(a^{(1)}\big)\lambda\pi\big(a^{(2)}\big)+\lambda\pi\big(a^{(1)}
\big)\lambda\pi\big(a^{(2)}\big)
\\
\hphantom{r_{\omega+\lambda}(a)}{}
=r_\omega(a)+D_\omega\lambda\pi(a)-\lambda\pi\big(a^{(2)}\big)\omega\pi\bigl(\kappa\big(a^{(1)}\big)a^{(3)}\bigr)
+\lambda\pi\big(a^{(1)}\big)\omega\pi\big(a^{(2)}\big)
\\
\hphantom{r_{\omega+\lambda}(a)=}{}
+\omega\pi\big(a^{(1)}\big)\lambda\pi\big(a^{(2)}\big)+\lambda\pi\big(a^{(1)}\big)\lambda\pi\big(a^{(2)}\big)
\\
\hphantom{r_{\omega+\lambda}(a) }{}
=r_\omega(a)+D_\omega\lambda\pi(a) +\lambda\pi\big(a^{(1)}\big)\lambda\pi\big(a^{(2)}\big)
\\
\hphantom{r_{\omega+\lambda}(a)=}{}
+\omega\pi\big(a^{(1)}\big)\lambda\pi\big(a^{(2)}\big)+\lambda\pi\big(a^{(3)}\big)\omega\bigl[\pi\big(a^{(1)}
\big)\circ\bigl(\kappa\big(a^{(2)}\big)a^{(4)}\bigr)\bigr]
\\
\hphantom{r_{\omega+\lambda}(a) }{}
=r_\omega(a)+D_\omega\lambda\pi(a)+\lambda\pi\big(a^{(1)}\big)\lambda\pi\big(a^{(2)}
\big)+\ell_\omega\bigl(\pi\big(a^{(1)}\big),\lambda\pi\big(a^{(2)}\big)\bigr).
\end{gather*}

There is the particular context of a~quantum principal bundle in which the structure group is interpreted
as a~`discrete object' and so the algebra of horizontal forms comes equipped with a~natural `zero
curvature' covariant derivative $D\colon\mathfrak{hor}(P)\rightarrow\mathfrak{hor}(P)$, which is actually
an arbitrary dif\/ferential.
In such a~context our starting point is a~graded dif\/ferential $*$-algebra, called $\mathfrak{hor}(P)$ say,
equipped with a~dif\/ferential $D$ and quantum group action
$F_\wedge\colon\mathfrak{hor}(P)\rightarrow\mathfrak{hor}(P)\otimes\mathcal{A}$ so that $F_\wedge
D=(D\otimes\mathrm{id})F_\wedge$.
In particular, $D$ satisf\/ies a~graded Leibniz rule and $D^2=0$.
In this case, we can construct a~natural complete calculus over $P$ as follows.
The graded $*$-algebra is given at the level of vector spaces by
$\Omega(P)=\mathfrak{hor}(P)\otimes\Gamma_{{\rm inv}}^\wedge$, while the product and $*$-structure are given
by
\begin{gather*}
(\varphi\otimes\vartheta)(\psi\otimes\eta)=(-)^{\partial\psi\partial\vartheta}\varphi\psi^{(0)}
\otimes\big(\vartheta\circ\psi^{(1)}\big)\eta
\qquad
\mathrm{and}\\
(\varphi\otimes\vartheta)^{*}=(-)^{\partial\varphi\partial\vartheta}\varphi^{(0)*}
\otimes\big(\vartheta^{*}\circ\varphi^{(1)*}\big).
\end{gather*}
There is a~canonical dif\/ferential $d_{\!P}\colon\Omega(P)\rightarrow\Omega(P)$ given by
\begin{gather*}
d_{\!P}(\varphi\otimes\vartheta)=D(\varphi)\otimes\vartheta+(-)^{\partial\varphi}\varphi^{(0)}
\otimes\pi\big(\varphi^{(1)}\big)\vartheta+(-)^{\partial\varphi}\varphi\otimes d^\wedge(\vartheta).
\end{gather*}
Here the dif\/ferential $d^\wedge\colon\Gamma^\wedge_{{\rm inv}}\rightarrow\Gamma^\wedge_{{\rm inv}}$ is
completely determined by its action on quantum germs and is given by
\begin{gather*}
d^\wedge\pi(a)=-\pi\big(a^{(1)}\big)\pi\big(a^{(2)}\big).
\end{gather*}
This can then be extended to the whole algebra $\Gamma_{{\rm inv}}^\wedge$.

The map $d_{\!P}\colon\Omega(P)\rightarrow\Omega(P)$ def\/ined by the above formula satisf\/ies
\begin{gather*}
d_{\!P}^2=0,
\qquad
d_{\!P}*=*d_{\!P},
\qquad
d_{\!P}(\mu\nu)=d_{\!P}(\mu)\nu+(-)^{\partial\mu}\mu d_{\!P}(\nu)
\end{gather*}
for all $\mu,\nu\in\Omega(P)$.

It is easy to verify that this is indeed a~dif\/ferential calculus on $P$ in the sense of our general
def\/inition.
In particular there exists a~canonical dif\/ferential extension of the co-action map to
$\widehat{F}\colon\Omega(P)\rightarrow\Omega(P)\mathbin{\widehat{\otimes}}\Gamma^\wedge$ which is
explicitly given by
\begin{gather*}
\widehat{F}(\varphi\otimes\Theta)=\varphi^{(0)}\otimes\Theta^{(1)}\otimes\varphi^{(1)}\Theta^{(2)},
\end{gather*}
where $\varphi\in\mathfrak{hor}(P)$ and $\Theta\in\Gamma^\wedge_{{\rm inv}}$.
Here we have put $\varphi^{(0)}\otimes\varphi^{(1)}=F_\wedge(\varphi)$ as well as
$\Theta^{(1)}\otimes\Theta^{(2)}=\widehat{\phi}(\Theta)$; let us also mention that
$\widehat{\phi}(\Gamma_{{\rm inv}}^\wedge)\subseteq\Gamma_{{\rm
inv}}^\wedge\mathbin{\widehat{\otimes}}\Gamma^\wedge$.

Let us also observe that we have a~distinguished connection $\varpi\colon\Gamma_{{\rm
inv}}\rightarrow\Omega^1(P)$ given by
\begin{gather*}
\varpi(\theta)=1\otimes\theta.
\end{gather*}

The covariant derivative of this connection is simply
\begin{gather*}
D_\varpi(\varphi)=d_{\!P}(\varphi)-(-)^{\partial\varphi}\varphi^{(0)}\otimes\pi\big(\varphi^{(1)}\big)=D(\varphi).
\end{gather*}

This calculus has been used as a~natural framework for Dunkl operators in the main part of this paper,
where $P$ is simply an open subset of a~f\/inite-dimensional Euclidean space~$E$ and~$G$ is a~Coxeter group
acting on this space and associated to a~root system ${R}$.

\section{Dif\/ferential calculi on f\/inite groups}\label{appendixB}

In this appendix we shall assume that $G$ is any f\/inite group.
We are going to describe all bicovariant $*$-covariant calculi over $G$.

Since in classical dif\/ferential geometry a~f\/inite group is a~zero-dimensional manifold (whose
inf\/initesimal structure is accordingly encoded in its zero-dimensional tangent spaces and so is trivial),
it is quite remarkable that in quantum group theory there exist non-trivial dif\/ferential (that is
`inf\/initesimal') calculi for any classical f\/inite group.
However, there is always more than one such dif\/ferential calculus (unless the group is trivial).

For the purposes of our paper, the principal interest will be when $G$ is a~f\/inite Coxeter group
(see~\cite{GB,HU}). However the main formulas here apply for all f\/inite groups.
See~\cite{D3} for the example $S_3$, the permutation group on $3$ letters.
This example can be easily generalized by the reader to $S_n$, this being a~Coxeter group.

We give $G$ the discrete topology (i.e., all subsets are open), since this is the only topology on $G$ that
is Hausdorf\/f.
Also, $G$ is compact with this topology.
So this is a~simple case of the Gelfand--Naimark theory, where in general one associates to a~compact,
Hausdorf\/f space its (dual) $C^{*}$-algebra of continuous, complex-valued functions.
In this case, since every function on~$G$ is continuous with respect to the discrete topology, the algebra
we get is
\begin{gather*}
\mathcal{A}=\{\beta\,|\,\beta:G\to\mathbb{C},
\
\mathrm{where\,}\beta\mathrm{\,is\,an\,arbitrary\,function}\}.
\end{gather*}
So the algebra $\mathcal{A}$ is commutative (with respect to pointwise multiplication of functions) and
f\/inite-dimensional, its dimension being the number of elements in $G$.

The elements of the group $G$ in a~natural way label a~basis in $\mathcal{A}$ by associating to every $g\in
G$ the Kronecker delta function whose support is the subset $\{g\}$ of $G$.
We denote this Kronecker delta by $g$.
So for $g,q\in G$ we have
\begin{gather*}
\sum_{g\in G}g=1,
\qquad
q\cdot g=
\begin{cases}
q,&\hbox{when~}q=g,
\\
0,&\hbox{otherwise.}
\end{cases}
\end{gather*}
In the above formula, the symbol $\cdot$ is used to denote the product of the Kronecker delta functions
that correspond to the elements $q,g\in G$, as already described.
The coproduct~$\phi$ and antipode~$\kappa$ act on these basis elements respectively by
\begin{gather*}
\phi(g)=\sum_{h\in G}h\otimes\big(h^{-1}g\big)
\qquad
\mathrm{and}
\qquad
\kappa(g)=g^{-1}.
\end{gather*}
These formulas follow from the general def\/inition that $\phi$ and $\kappa$ are the pull-backs to
functions (i.e., elements of $\mathcal{A}$) of the group multiplication $G\times G\to G$ ($(g,h)\mapsto
gh$) and group inversion $G\to G$ ($g\mapsto g^{-1}$), respectively.
We also have the following expression
\begin{gather*}
\mathrm{ad}(g)=\sum_{h\in G}\big(hgh^{-1}\big)\otimes h
\end{gather*}
for the adjoint action.
The elements $g\in G$ also can be interpreted via the map of evaluation at~$g$ as characters on
$\mathcal{A}$: For all $\beta\in\mathcal{A}$ we can write $g(\beta)=\beta(g)$.
In particular, the counit map $\epsilon\colon\mathcal{A}\to\mathbb{C}$, being a~character, corresponds to
the neutral element $\epsilon\in G$, and so this justif\/ies our using the same symbol $\epsilon$ for them.

The ideals in $\mathcal{A}$ are all $*$-ideals and are naturally labeled by subsets of~$G$, consisting of all
points where all the elements of the ideal vanish.
In the theory of dif\/ferential calculi, we are interested in the ideals $\mathcal{R}$ satisfying
$\mathcal{R}\subseteq\ker(\epsilon)$.
Such ideals correspond to subsets of~$G$ containing the neutral element~$\epsilon$.
Hence there is a~natural correspondence between these ideals~$\mathcal{R}$ and subsets~$S$ of
$G\setminus\{\epsilon\}$.
We can write ${\mathcal{R}}={\mathcal{R}}_S$, for such ideals ${\mathcal{R}}$ and $S=S_{{\mathcal{R}}}$ for
such subsets.
To put it in terms of precise expressions we have
\begin{gather*}
{\mathcal{R}}_S=\bigl\{f\in\ker(\epsilon)\big \vert f(x)=0
\
 \forall\,  x\in S\bigr\},
\qquad
S_{{\mathcal{R}}}=\bigl\{x\in G\setminus\{\epsilon\}\big\vert f(x)=0
\
 \forall\,  f\in{\mathcal{R}}_S\bigr\}.
\end{gather*}

Because of the natural isomorphism $\Gamma_{{\rm inv}}\leftrightarrow\ker(\epsilon)/R$, as explained
in~\cite{W-diff} and~\cite{D1}, there is a~canonical basis in the space $\Gamma_{{\rm inv}}$ given by the
elements $\bigl\{\pi(g)=[g]\bigm\vert g\in S_{{\mathcal{R}}}\bigr\}$.
The right $\mathcal{A}$-module structure $\circ$ is specif\/ied by
\begin{gather*}
[g]\circ h=
\begin{cases}
[g],&\text{if}~g=h,
\\
0, &\text{otherwise},
\end{cases}
\end{gather*}
and the complete right $\mathcal{A}$-module structure on the calculus $\Gamma$ is def\/ined by
\begin{gather*}
[g]h=\big(hg^{-1}\big)[g].
\end{gather*}

It is worth mentioning that
\begin{gather}
\label{e-germ}
[\epsilon]=-\sum_{s\in S}[s],
\end{gather}
as follows from the fact that elements $g\in G$ sum up to $1$, and that the germs map $\pi$ acts
non-trivially on $S$ and $\epsilon$ only.
The property of $*$-covariance $\kappa({\mathcal{R}})^{*}={\mathcal{R}}$ is equivalent to
$S_{{\mathcal{R}}}^{-1}=S_{{\mathcal{R}}}$ or, in other words, the set $S$ should contain with every
element $s$ its inverse $s^{-1}$.

On the other hand, the bicovariance property
$\mathrm{ad}({\mathcal{R}})\subseteq{\mathcal{R}}\otimes\mathcal{A}$ is equivalent to
$gS_{{\mathcal{R}}}g^{-1}=S_{{\mathcal{R}}}$ for all $g\in G$ or, in other words, the set
$S_{{\mathcal{R}}}$ is invariant under the adjoint action of $G$.
Equivalently~$S_{{\mathcal{R}}}$ can be written as a~(disjoint) union of conjugation classes.

In particular, we see that irreducible $*$-covariant bicovariant calculi are in a~natural correspondence with
the subsets of $G$ which are either \bla{i} a~single conjugate class that coincides with its inverse or
\bla{ii} a~union of two distinct mutually inverse conjugation classes.
The notion of irreducibility is understood as the non-existence of another non-trivial dif\/ferential
$*$-covariant and bicovariant calculus onto which the given calculus can be projected.

The canonical braid-operator $\sigma\colon\Gamma_{{\rm inv}}^{\otimes2}\rightarrow\Gamma_{{\rm
inv}}^{\otimes2}$ of~\cite{W-diff} can be calculated to be
\begin{gather*}
\sigma\bigl([h]\otimes[g]\bigr)=\big[hgh^{-1}\big]\otimes[h].
\end{gather*}
We have in particular that if $gh=hg$, then $\sigma$ acts as the standard f\/lip.
The elements $[q]\otimes[q]$ are always $\sigma$-symmetric.
In general, this operator will have eigenvalues which are complex roots of unity, since the above formula
gives us a~bijective map from the f\/inite set $S\times S$ to itself, in other words
\begin{gather*}
\sigma\colon \ (h,g)\mapsto\big(hgh^{-1},h\big).
\end{gather*}
So the orbits of this action are used to construct the eigenvectors of $\sigma$.
Note that the diagonal elements $[q]\times[q]$ are the only one-element orbits, and the two-element orbits
are precisely given by $\{(h,g),(g,h)\}$ where $g$ and $h$ are dif\/ferent elements of $S_{{\mathcal{R}}}$
that commute.
Each orbit naturally generates the space of a~regular representation of a~f\/inite cyclic group whose order
is equal to the number of elements in that orbit.

In terms of all these identif\/ications, the universal dif\/ferential envelope $\Gamma^\wedge$ of $\Gamma$
is constructible in terms of the following quadratic relations:
\begin{gather*}
\sum_{gq=h}[g]\otimes[q],
\qquad
\epsilon\neq h\not\in S_{{\mathcal{R}}},
\qquad
q,g\in S_{{\mathcal{R}}}.
\end{gather*}
In particular, we see that the dimension of the quadratic relations space in $\Gamma_{{\rm
inv}}\otimes\Gamma_{{\rm inv}}$ is the same as the cardinality of the set of non-identity elements of
$G\setminus S_{{\mathcal{R}}}$ which can be expressed as a~product of two elements from $S_{{\mathcal{R}}}$.

Let us now apply all this in the context of quantum principal bundles and maps
$T_\lambda\colon\mathfrak{hor}(P)\rightarrow\mathfrak{hor}(P)$ induced by connection displacements
$\lambda$.
Such maps naturally appear in the expressions for covariant derivatives and are given by
\begin{gather*}
T_\lambda(\varphi)=\varphi^{(0)}\lambda\pi\big(\varphi^{(1)}\big).
\end{gather*}
Taking into account the structural form of the calculus and in particular the identity~\eqref{e-germ} we
obtain
\begin{gather*}
T_\lambda(\varphi)=\sum_{s\in S}(\varphi_s-\varphi)\lambda[s],
\end{gather*}
where we used an alternative form to describe the action $F_\wedge$ via
\begin{gather}\label{Fw-G}
F_\wedge(\varphi)=\varphi^{(0)}\otimes\varphi^{(1)}=\sum_{g\in G}\varphi_g\otimes g,
\end{gather}
so that
\begin{gather*}
\varphi_g=(\mathrm{id}\otimes g)F_\wedge(\varphi).
\end{gather*}
Here $g\in G$ is interpreted as a~character $g\colon\mathcal{A}\rightarrow\mathbb{C}$ via evaluation as
described earlier.

The displaced covariant derivative is thus given by
\begin{gather*}
D_{\omega+\lambda}(\varphi)=D_\omega(\varphi)+(-)^{\partial\varphi}\sum_{s\in S}
(\varphi-\varphi_s)\lambda[s].
\end{gather*}
The displacement map $\lambda$ is $*$-invariant, which translates into
\begin{gather}
\label{lambda-*}
(\lambda[s])^{*}=-\lambda\big[s^{-1}\big].
\end{gather}
It is also $\mathrm{ad}$-covariant, which means that
\begin{gather*}
F_\wedge\lambda[s]=\sum_{g\in G}(\lambda[s])_g\otimes g=\lambda\big[s^{(2)}\big]\otimes\kappa\big(s^{(1)}\big)s^{(3)}
=\sum_{g\in G}\lambda\big[gsg^{-1}\big]\otimes g.
\end{gather*}
This can be simply expressed as
\begin{gather}
\label{lambda-covariance}
\lambda[s]_g=\lambda\big[gsg^{-1}\big]
\end{gather}
for every $s\in S$ and $g\in G$.

Now for an irreducible, bicovariant and $*$-covariant calculus $\Gamma$ these properties tell us that the
displacement $\lambda$ is completely determined by the value $\lambda[s]\in\mathfrak{hor}^1(P)$ on a~single
element $[s]$ where $s\in S$.
In the case when the calculus is based on the union of two disjoint conjugation classes (each being the
inverse of the other), we can take $\lambda[s]$ to be an arbitrary horizontal one-form and then
use~\eqref{lambda-*} and~\eqref{lambda-covariance} to def\/ine $\lambda$ on all of $\Gamma_{{\rm inv}}$.

On the other hand, when there is only one conjugation class (being its own inverse) used to def\/ine the
calculus, we have
\begin{gather*}
\lambda[s]^{*}=-\lambda[s]_q,
\end{gather*}
where $q\in G$ is such that $s^{-1}=qsq^{-1}$.
The same applies in a~more general context, for arbitrary elements of $S$ conjugated to their inverse.

In the general case of an arbitrary bicovariant and $*$-covariant calculus $\Gamma$ the above reasoning
should be repeated for all the irreducible blocks.

Let us now consider an important special case of the above situation, when the algebra $\mathfrak{hor}(P)$
admits a~`local trivialization'.
This means that we can establish in a~covariant way projection maps of the form
\begin{gather*}
\mathfrak{hor}(P)\rightsquigarrow\mathcal{L}\otimes\mathcal{A},
\end{gather*}
where $\mathcal{L}$ is a~`local representation' of the calculus on the base space $M$.
For example, we can consider a~completely classical horizontal forms algebra, where $M$ and $P$ are
classical, and $\mathfrak{hor}(P)$ is the classical algebra of dif\/ferential forms on $P$.

Then locally the map $\lambda$ is given by
\begin{gather*}
\lambda[s]\leftrightarrow Y\big[s^{(2)}\big]\otimes\kappa\big(s^{(1)}\big)s^{(3)}=\sum_{g\in G}Y\big[gsg^{-1}\big]\otimes g,
\end{gather*}
where $Y\colon\Gamma_{{\rm inv}}\rightarrow\mathcal{L}$ is a~linear map.
In this form the covariance condition is automatically satisf\/ied.
The $*$-condition ref\/lects as an appropriate restriction on $Y$.
For example, if the local trivialization intertwines the $*$-structure on $\mathfrak{hor}(P)$ and the product
of $*$-structures on $\mathcal{L}$ and $\mathcal{A}$, then $(Y[s])^{*}=-Y[s^{-1}]$.

We have covered the main properties of the dif\/ferential calculi and connections associated to f\/inite
groups $G$, interpreted as structure groups of quantum principal bundles.
For the purposes of this paper we are interested in the special case when $G$ is a~Coxeter group.
Specif\/ic formulas and considerations for that case are collected in Section~\ref{results}.

\subsection*{Acknowledgments}

The second author wishes to thank the Instituto de Matem\'aticas (UNAM) and the f\/irst \mbox{author} for their
generous hospitality during various academic visits during which this paper was written.
The f\/irst author would like to express his gratitude to the Centro de Investigaciones en Matem\'aticas
(CIMAT, Guanajuato) and the second author for their kind hospitality during several academic visits during
which the roots of the conceptual framework for this research were established.
We both gratefully thank the referees for their comments which have led to several clarif\/ications and
improvements.

\pdfbookmark[1]{References}{ref}
\LastPageEnding

\end{document}